\documentclass[12pt,a4]{article}

\usepackage{amsmath,amsthm,amssymb}
\usepackage{a4wide}
\usepackage[dvips]{graphicx}

\usepackage{mybbold}    

\usepackage[T1]{fontenc}
\usepackage[bf,small]{caption2}

\usepackage{def}

\begin{document}

\thispagestyle{empty}

\noindent
December 2002\\

\vspace*{1cm}

\begin{center}

{\LARGE\bf 
Semiclassical quantisation rules for the Dirac and Pauli equations} \\
\vspace{2cm}
{\large \bf Stefan Keppeler}

\vspace{1cm}
\parbox[t]{5.5cm}{
Abteilung Theoretische Physik\\
Universit\"at Ulm\\
Albert-Einstein-Allee 11\\ 
D-89069 Ulm, Germany }
\hspace{2cm}
\parbox[t]{5cm}{
Matematisk Fysik\\
Lunds Tekniska H\"ogskola\\
Lunds Universitet, Box 118\\
SE-22100 Lund, Sweden}%
\footnote{present address, email: {\tt stefan.keppeler@matfys.lth.se}}%
\end{center}

\vfill

\begin{abstract}
We derive explicit semiclassical quantisation conditions for the Dirac 
and Pauli equations. We show that the spin degree of freedom yields 
a contribution which is of the same order of magnitude as the Maslov 
correction in Einstein-Brillouin-Keller quantisation. In order to obtain
this result a generalisation of the notion of integrability for a 
certain skew product flow of classical translational dynamics and classical 
spin  precession has to be derived. Among the examples discussed is the 
relativistic Kepler problem with Thomas precession, whose treatment sheds
some light on the amazing success of Sommerfeld's theory of fine structure
[Ann. Phys. (Leipzig) {\bf 51} (1916) 1--91].
\end{abstract}

\newpage

\section{Introduction}

Semiclassical quantisation conditions provide the most direct link between 
the old quantum theory of Bohr and Sommerfeld on the one hand and wave 
mechanics on the other hand. Unlike other semiclassical tools, like trace 
formulae or the Van Vleck-Gutzwiller propagator \cite{Gut90}, 
they do not only express 
quantum mechanical objects in terms of classical properties but also  employ
exactly the same formulation as was used by the old quantum theory, namely
action quantisation.

Before the advent of quantum mechanics quantisation of a system was done by
determining action integrals within the classical theory and setting these 
equal to an integer multiple of Planck's constant $h=2\pi\hbar$, 
i.e. one required
\begin{equation}
\label{bohr}
  \oint p \, \ud x = 2\pi\hbar \, n 
\end{equation}
with integer $n$. This condition, originally put forward by Bohr \cite{Boh13} 
in order to understand the hydrogen spectrum, was first understood as the 
quantisation rule for one degree of freedom. 

Around 1915 there was an ongoing discussion 
how this condition should be translated to more than one degree of freedom, 
see e.g. \cite{Pla16,Eps16,Sch16} and the introduction of \cite{Som16}. 
Epstein \cite{Eps16} proposed to use that set of coordinates in which the 
problem separates if such coordinates exist. For each degree of freedom there 
would then be a condition of the form \eqref{bohr}, i.e.
\begin{equation}
\label{epstein}
 \oint p_j \, \ud x_j = 2\pi\hbar \, n_j 
\end{equation}
with integers $n_j$, $j$ numbering the degrees of freedom and $(p_j,x_j)$ 
being a pair of canonically conjugate variables in that particular set of 
coordinates. Epstein's point of view was assumed by 
Sommerfeld who successfully applied this prescription treating many problems
in spectroscopy \cite{Som16}. 

Shortly after, Einstein \cite{Ein17}  pointed out that separability of 
the equations of motion is 
not a necessary condition for action quantisation but that merely 
integrability (in the sense of Liouville and Arnold \cite{Lio55,Arn78})
is required: If there are sufficiently many integrals of motion with 
pairwise commuting Poisson brackets then the phase space foliates into 
invariant tori on which the line integral 
\begin{equation}
  \int_{\vecx_{\rm i}}^{\vecx_{\rm f}} \vecp \, \ud\vecx
\end{equation}
is locally path-independent. The quantisation conditions can then be written 
in the form 
\begin{equation}
\label{einstein}
  \oint_{\mathcal{C}_j} \vecp \, \ud\vecx = 2\pi\hbar \, n _j 
\end{equation}
where $\{ \mathcal{C}_1, \hdots, \mathcal{C}_d \}$ denotes a basis of
non-contractible loops on a given torus. This formulation has the advantage 
over Epstein's version of being independent of the coordinate system. 

Soon after the old quantum theory had been replaced by matrix and wave 
mechanics the old quantisation conditions were rederived and modified by 
Wentzel \cite{Wen26}, Kramers \cite{Kra26}, Brillouin \cite{Bri26} and 
Jeffreys \cite{Jef25} in a short-wavelength approximation, the so-called
WKB or JWKB method. It was shown that depending on the character of the motion 
the quantum numbers may have to be shifted by a small number, leading e.g. 
to half-integer quantum numbers for oscillations but integer quantum numbers
for rotations. Again the treatment was first for one-dimensional 
and then for separable systems. A complete derivation 
of the quantisation conditions from the Schr\"odinger equation that takes 
into account both the abstract integrability condition used by Einstein
and the small shift of the quantum numbers is due to Keller \cite{Kel58}.
He proved the semiclassical quantisation conditions 
\begin{equation}
\label{EBK}
  \oint_{\mathcal{C}_j} \vecp \, \ud \vecx 
  = 2\pi\hbar \left( n_j + \frac{\mu_j}{4} \right)
\end{equation}
which are now known as Einstein-Brillouin-Keller (EBK) or torus 
quantisation. The number $\mu_j \in \Z_4$ denotes the Maslov index,
see e.g. \cite{Mas72,MasFed81}, a topological invariant of the cycle 
$\mathcal{C}_j$. In one dimension it counts the number of turning points 
encountered along the loop. 
A good overview on theses topics is given in \cite{Kel85}.

The discussion so far was only for non-relativistic quantum mechanics, i.e 
the semiclassical approximation for the Schr\"odinger equation.
In 1916 Sommerfeld applied the quantisation conditions \eqref{epstein}
also in a relativistic context \cite{Som16}. 
His aim was to find small corrections to 
the hydrogen spectrum which he expected to be due to relativistic effects.
The success was overwhelming, the so-called Sommerfeld fine structure formula 
agreed excellently with the experimental data. More than ten years later 
\cite{Gor28,Dar28} it was found that the energy levels of the hydrogen 
atom, when calculated using the Dirac equation, the correct relativistic 
wave equation for the electron, are identical to the levels determined by 
Sommerfeld. The Dirac equation, however, does not only take into account 
relativistic effects but also the half-integer spin of the electron. 
The fine structure accounted for by the Sommerfeld formula is to a large 
extent due to spin-orbit coupling, an effect that was unknown at the 
time Sommerfeld did his calculations. In fact, even the property of spin 
itself was yet to be discovered. This seeming paradox has to be explained
by a semiclassical analysis of the Dirac equation.

Early semiclassical approaches to the Dirac equation are due to 
Pauli \cite{Pau32} and Rubinow and Keller \cite{RubKel63}. These will 
be discussed in section \ref{sec:scDirac}. The subtleties connected with 
such an approach are related to the fact that the Dirac equation is a
partial differential equation for a spinor and not just for a scalar 
wave function.
Therefore these problems should be discussed in the more general context of 
semiclassical (or short wavelength) approximations to multicomponent 
wave equations. It was observed by Yabana and Horiuchi \cite{YabHor86}
that the occurrence of geometrical or Berry phases \cite{Ber84,ShaWil89}
plays an important r\^ole in this context. Kuratsuji and Iida 
\cite{KurIid85,KurIid88}, using path 
integral methods, suggested that the symplectic structure of phase space 
should be deformed such that it includes the contribution of geometric 
phases. A review of these results is given in \cite{LitFly91b}. 
A general method for the semiclassical quantisation of multi-component 
wave equation was derived by Littlejohn and Flynn \cite{LitFly91a,LitFly91b}.
Their results hold whenever the principal Weyl symbol of the Hamiltonian, 
which is a matrix valued function on classical phase, has non-degenerate 
eigenvalues. Thus it does not apply to the Dirac equation, in which case
these eigenvalues have a multiplicity of two as will be shown in section 
\ref{sec:scDirac}. In such a situation the geometrical phases that appear 
in the semiclassical expressions are not simple $\U(1)$-phases any longer but 
are themselves matrix-valued and thus in general non-Abelian. It was shown 
by Emmrich and Weinstein \cite{EmmWei96} that in this case  
integrability of the dynamics generated by the classical ray Hamiltonians 
(the eigenvalues of the principal Weyl symbol) is no longer sufficient 
to guarantee the existence of global semiclassical wave functions and thus
of semiclassical quantisation conditions.

In this article we derive semiclassical quantisation conditions 
from the Dirac equation, the relativistic wave equation for particles 
with spin $\frac{1}{2}$, and from the Pauli equation, describing 
particles with arbitrary spin in a non-relativistic context.
We show how to resolve the problems mentioned in the context with the 
occurrence of non-Abelian Berry phases by developing a generalisation 
of the notion of integrability that not only includes the dynamics of the 
ray Hamiltonians but imposes an additional condition. In this way we 
effectively reduce the non-Abelian phases to $\U(1)$-phases. The latter 
enter the semiclassical quantisation conditions by a correction which 
is of the same order as the Maslov contribution in \eqref{EBK}.
This correction represents the influence of the spin degree of 
freedom and can be given a clear physical interpretation in terms of
classical spin precession.
By applying the method to the relativistic Kepler problem we 
shed some light on the success of Sommerfeld's fine structure formula.
A brief account of some of these results was given in \cite{Kep02}.

The paper is organised as follows. In section \ref{sec_ebk} we outline 
the derivation of EBK-quantisation for later reference, thereby 
emphasising the r\^ole of integrability. Section 
\ref{sec:scDirac} deals with the determination of semiclassical wave 
functions for the Dirac equation. In section \ref{sec:integrable_skew} 
we generalise the  concept of integrability to the case of group
extensions and the skew product of classical translational dynamics
and classical spin precession. Based on this characterisation we then 
derive explicit semiclassical quantisation conditions for the Dirac 
equation in section \ref{sec:rotation_angels}. In section \ref{sec:pauli}
we show how these results translate to the Pauli equation with arbitrary 
spin. Before we treat some special examples in sections 
\ref{sec:oscillator} and \ref{sec:sommerfeld} (among which is Sommerfeld's 
theory of fine structure) we derive general formulae which facilitate the
semiclassical quantisation of spherically symmetric systems in section  
\ref{sec:spherical}. We conclude with a summary in section 
\ref{sec:conclusions}. Some important formulae for Weyl quantisation are 
listed in appendix \ref{weyl}.

\section{EBK quantisation}
\label{sec_ebk}

In this section we briefly summarise the main steps in the derivation of 
the EBK quantisation rules. We do so in order to introduce some notation  
and for later reference such that we can 
explicitly compare to this basic situation when treating systems with spin.
This is, however, not intended to be a complete review of EBK quantisation 
and we thus refer the reader seeking a comprehensive introduction to EBK
quantisation to the literature, e.g. \cite{Kel58,Kel85}.

We want to find an asymptotic solution of the stationary 
Schr\"odinger equation, 
\begin{equation}
\label{stationary_Schroedinger}
  \op{H} \psi(\vecx) = E \psi(\vecx) \, , 
\end{equation}
where $\psi \in L^2(\R^d)$ and $\op{H}$ shall be a Weyl operator
(some facts on Weyl quantisation are summarised in appendix \ref{weyl})
with symbol 
\begin{equation}
  H(\vecp,\vecx) 
  = H_0(\vecp,\vecx) + \hbar H_1(\vecp,\vecx) + \O(\hbar^2) \, , \quad
  \hbar \to 0 \, .
\end{equation}
The leading order term $H_0$ in the semiclassical limit $\hbar \to 0$
is known as the principal symbol of $\op{H}$ and $H_1$ is called 
the sub-principal symbol. For simplicity we will choose $H_1 \equiv 0$ 
for the rest of this section. For the familiar Hamiltonian 
\begin{equation}
\label{HamOp_mit_V}
  \op{H} = -\frac{\hbar^2}{2m} \Delta + V(\vecx)
\end{equation}
describing a particle of mass $m$ moving under the influence of the external 
potential $V(\vecx)$ the Weyl symbol reads
\begin{equation}
  H(\vecp,\vecx) \equiv H_0(\vecp,\vecx) = \frac{\vecp^2}{2m} + V(\vecx)
\end{equation}
and we can write $\op{H}=H(\frac{\hbar}{\ui}\nabla,\vecx)$.

For the wave function $\psi$ one tries the WKB ansatz 
\begin{equation}
\label{WKBansatz}
  \psi_\mathrm{WKB}(\vecx) = 
  \sum_{k\geq0}\left(\frac{\hbar}{\ui}\right)^k \, a_k(\vecx) \, 
  \ue^{\frac{\ui}{\hbar} S(\vecx)} \, .
\end{equation} 
Inserting into \eqref{stationary_Schroedinger} yields in leading 
orders in $\hbar$, cf. appendix \ref{weyl},
\begin{equation}
\label{insertWKBansatz}
\begin{split}
  &\left[ H(\nab{x}S,\vecx) - E \right] a_0 
  + \frac{\hbar}{\ui} \Big\{ \left[ H(\nab{x}S,\vecx) - E \right] a_1 \\
  &+ \left[ (\nab{p}H)(\nab{x}S,\vecx) \right](\nab{x}a_0) 
  + \frac{1}{2} \left[ \nab{x} (\nab{p}H)(\nab{x}S,\vecx) \right] a_0 \Big\} 
  + \O(\hbar^2) = 0 \, .
\end{split}
\end{equation}
This equation is easily confirmed for the particular Hamiltonian 
\eqref{HamOp_mit_V} by direct computation, but it also holds for arbitrary 
(semiclassical) Weyl operators, see appendix \ref{weyl}. The strategy of the 
WKB method is now to satisfy eq. \eqref{insertWKBansatz} separately order
by order in $\hbar$. In leading order one finds the Hamilton-Jacobi equation 
\begin{equation}
\label{HJG}
  H(\nab{x}S,\vecx) = E 
\end{equation}
of classical mechanics with the (principal) symbol $H(\vecp,\vecx)$ acting as 
the classical Hamiltonian. From standard Hamilton-Jacobi theory, 
see e.g. \cite{Gol80,Arn78}, one thus concludes that the phase $S(\vecx)$ 
of the WKB ansatz becomes the classical action generating the dynamics 
with Hamiltonian $H(\vecp,\vecx)$: If $(\vecP(t),\vecX(t))$ is a solution 
of Hamilton's equations of motion then $\nab{x}S(\vecX(t))=\vecP(t)$.
A solution of the Hamilton-Jacobi equation \eqref{HJG} can thus be obtained by
integration along solutions of Hamilton's equations of motion in the
following way. Denote by $\vecy$ an arbitrary point in configuration space
and by $\vecxi$ a momentum satisfying $H(\vecxi,\vecy)=E$. Let $\phi_H^t$ be
the (Hamiltonian) flow generated by the classical Hamiltonian $H(\vecp,\vecx)$,
i.e. $\phi_H^t(\vecxi,\vecy) =: (\vecP(t),\vecX(t))$ 
denotes the point reached at time $t$ 
on the trajectory starting at $(\vecxi,\vecy)$. Then we have 
\begin{equation}
  S(\vecx)-S(\vecy) 
  = \int_0^t \left[ \frac{\ud}{\ud t^\prime} S(\vecX(t^\prime)) \right]
    \ud t^\prime
  = \int_0^t \nab{x}S(\vecX(t^\prime)) \, \dot{\vecX}(t^\prime) \, \ud t^\prime
  = \int_{\vecy}^{\vecx} \vecP \, \ud \vecX
\end{equation}
where in the last expression integration is along the trajectory 
$\phi_H^t(\vecxi,\vecy)$. Finally the action reads
\begin{equation}
\label{S_integrated}
  S(\vecx) = S(\vecy) + \int_{\vecy}^{\vecx} \vecP \, \ud \vecX  \, , 
\end{equation}
where $S(\vecy)$ is the arbitrarily chosen value of $S$ at the point $\vecy$.
Given a solution \eqref{S_integrated} of the Hamilton-Jacobi equation 
the next-to-leading order equation deriving from \eqref{insertWKBansatz}
reduces to 
\begin{equation}
\label{scalar_transport}
  \left[ (\nab{p}H)(\nab{x}S,\vecx) \right](\nab{x}a_0) 
  + \frac{1}{2} \left[ \nab{x} (\nab{p}H)(\nab{x}S,\vecx) \right] a_0
  = 0 \, .
\end{equation}
This is known as the transport equation for the leading order 
amplitude $a_0(\vecx)$.
Due to Hamilton's equations of motion the first term can now
be interpreted as a time derivative along the the trajectory 
$\phi_H^t(\vecxi,\vecy)$ which we shall denote by $\frac{\ud}{\ud t}$ 
or simply by a dot, 
\begin{equation}
\label{def_dot}
  \dot{a} \equiv \frac{\ud a}{\ud t} := 
  \left[ (\nab{p}H)(\nab{x}S,\vecx) \right](\nab{x}a) \, .
\end{equation}
The solution of \eqref{scalar_transport} is locally given by
\begin{equation}
\label{scalar_a0}
  a_0(\vecx) = \sqrt{\det\frac{\partial \vecy}{\partial \vecx}} \, , 
\end{equation}
see e.g. \cite{Kel58,Kel85}. Together with \eqref{S_integrated} one has thus 
found an approximate solution
\begin{equation}
\label{WKBloes}
  \psi_\mathrm{WKB}(\vecx) \sim 
  \sqrt{\det\frac{\partial \vecy}{\partial \vecx}} \, 
  \exp\left( \frac{\ui}{\hbar}S(\vecy) 
       + \frac{\ui}{\hbar} \int_{\vecy}^{\vecx} \vecP \, \ud\vecX \right)
\end{equation}
of the Schr\"odinger equation for points $\vecx$ in a neighbourhood 
of $\vecy$ that are visited along a a solution of Hamilton's 
equations of motion starting at $\vecy$. 

However, one still needs to find 
a way to integrate the Hamilton-Jacobi equation along paths transversal 
to $\phi_H^t$. Furthermore, the approximation \eqref{WKBloes} breaks down at
points where $\phi_H^t(\vecxi,\vecy)$ touches a caustic and thus 
$\frac{\partial \vecy}{\partial \vecx}$ becomes singular. For one degree of
freedom these points are given by the turning points. In order to address
both these problems one has to know more about the classical phase 
space structure. For arbitrary Hamiltonians one can in general not proceed 
far beyond this point as was already pointed out by Einstein \cite{Ein17}
in the context of the old quantum theory. Instead one has to invoke the
concept of integrability. 

Following Liouville \cite{Lio55} we say that a Hamiltonian is integrable
if there are $d$ constants of motion $A_1:=H,A_2,\hdots,A_d$ which are 
in involution, i.e. whose Poisson brackets vanish pairwise.
A complete proof of the consequences of this definition is due to Arnold 
whose version \cite[chapter 10]{Arn78} we shall quote here.
\begin{theorem}(Liouville-Arnold)
\label{Liouville-Arnold}
Suppose that we are given $d$ functions in involution on a $d$-dimensional 
symplectic manifold
\begin{equation}
\label{involution}
  A_1,\hdots,A_d \, , \quad 
  \{ A_j,A_k\} \equiv 0 \, , \quad j,k = 1,\hdots,d \, .
\end{equation}
Consider a level set of the functions $A_j$,
\begin{equation}
  M_{\veca} = 
  \{ (\vecp,\vecx) \, | \ A_j(\vecp,\vecx) = a_j \, , \ j=1,\hdots,d \} \, .
\end{equation}
Assume that the $d$ functions $A_j$ are independent on $M_{\veca}$ (i.e. the
$d$ 1-forms $\ud A_j$ are linearly independent at each point of $M_{\veca}$). 
Then
\begin{enumerate}
\item $M_{\veca}$ is a smooth manifold, invariant under the 
      phase flow with Hamiltonian function $H=A_1$.
\item If the manifold $M_{\veca}$ is compact and connected, then it is 
      diffeomorphic to the $d$-dimensional torus
      \begin{equation}
        \T^d = \{ (\vartheta_j,\hdots,\vartheta_d) \mod \ 2\pi \} \, .
      \end{equation}
\item The phase flow with Hamiltonian function $H$ determines a conditionally 
      periodic motion on $M_{\veca}$, i.e. in angular coordinates 
      $\vecvt = (\vartheta_j,\hdots,\vartheta_d)$ we have 
      \begin{equation}
      \label{theta_punkt}
        \frac{\ud \vecvt}{\ud t} = \vecomega \, , \quad 
        \vecomega = \vecomega(\veca) \, .
      \end{equation}
\item The canonical equations with Hamiltonian function $H$ can be 
      integrated by quadratures.
\end{enumerate}
\end{theorem}
\noindent
We refrain from providing a proof of the theorem here but refer the reader 
to Arnold's book \cite{Arn78}. Instead we remark on some aspects which 
are relevant for the following sections. 

In order to prove property 2 one first shows that the conditions 
\eqref{involution} imply that the flows $\phi_j^t$ generated by the 
observables $A_j$, $j=1,\hdots,d$, commute on $M_{\veca}$, i.e.
\begin{equation}
\label{phis_commute}
  \phi_j^t \circ \phi_k^{t^\prime} = \phi_k^{t^\prime} \circ \phi_j^t 
  \quad \forall \ j,k=1,\hdots,d \, .
\end{equation}
this yields a transitive action of $\R^d$ on $M_{\veca}$. In the following 
we will refer to $M_{\veca}$ as a Liouville-Arnold torus.

In general the coordinates $\vecvt$ and the observables $\vecA$ are not
canonically conjugate. However, (locally) there exists a mapping 
$\vecA \mapsto \vecI$ such that $(\vecI,\vecvt)$ form a set of canonically
conjugate variables. The new constants of motion $\vecI(\vecA)$ are 
called action variables and the explicit construction of action and angle 
variables $(\vecI,\vecvt)$ is the desired integration of of Hamilton's 
equations of motion by quadratures. 
The Hamiltonian expressed in the new variables 
becomes a function $\overline{H}(\vecI)$ of the action variables only and
thus the frequencies in \eqref{theta_punkt} are given by 
$\vecomega = \nab{I}\overline{H}(\vecI)$.

Theorem \ref{Liouville-Arnold} can be used in order to derive the EBK 
quantisation conditions as follows. Since the flows $\phi_j^t$ commute we 
can define the action $S(\vecx)$ analogously to \eqref{S_integrated} by 
integration along the flow lines of $\phi_2^t,\hdots,\phi_d^t$ instead of 
$\phi_1^t \equiv \phi_H^t$. To this end define the multi-time flow 
\begin{equation}
  \Phi^{\vect} := \phi_d^{t_d} \circ \cdots \circ \phi_1^{t_1}  \, , 
\end{equation}
where due to \eqref{phis_commute} the ordering is unimportant. Since 
$\Phi^{\vect}$ is a transitive action of $\R^d$ on a Liouville-Arnold torus,
for any $\vecx$ in a small neighbourhood of $\vecy$ there is 
a unique $\vect\in\R^d$ such that 
\begin{equation}
  \Phi^{\vect}(\vecxi,\vecy) = (\vecp,\vecx)
\end{equation}
with some momentum $\vecp$. The rapidly oscillating phase of the WKB 
wave function is then given by 
\begin{equation}
  S(\vecx) = S(\vecy) + \int_{\vecy}^{\vecx} \vecP \, \ud\vecX
\end{equation}
where integration is along $\Phi^{\vect}(\vecxi,\vecy)$. Since the flows 
$\phi_j^t$, $j=1,\hdots,d$, commute this is not in conflict with the 
requirement of $S$ solving the Hamilton-Jacobi equation \eqref{HJG}.
This answers the question of how to integrate transversal to $\phi_H^t$. 

\begin{figure}[t]
\begin{center}
\includegraphics[width=15cm]{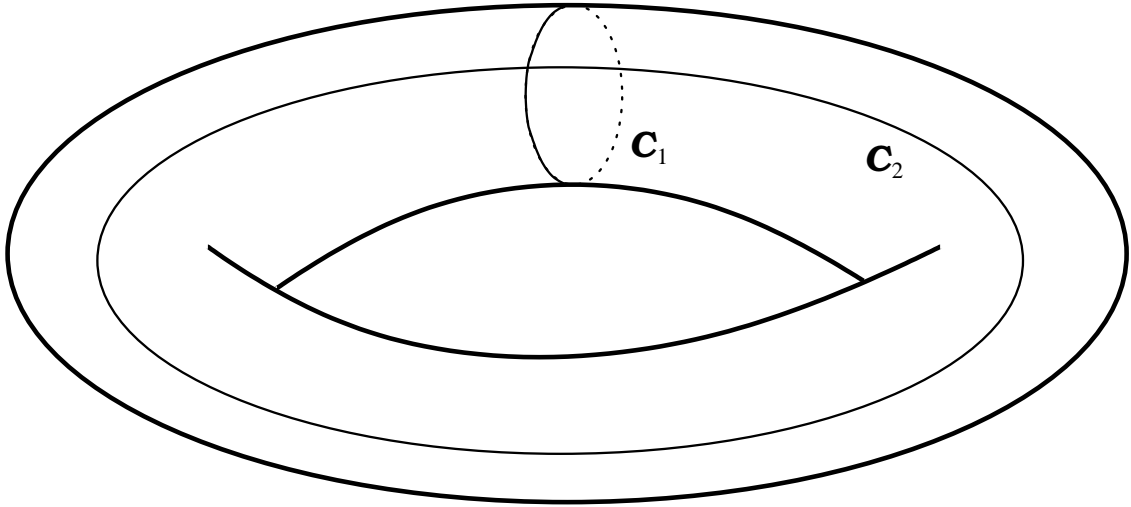}
\caption{Sketch of a basis $\{\mathcal{C}_1,\mathcal{C}_2\}$ of loops on a 
two-torus $\T^2$.} 
\label{fig:torus}
\end{center}
\end{figure}

One still has to solve the problem of 
$a_0=\sqrt{\det\frac{\partial\vecy}{\partial\vecx}}$ becoming singular 
for certain values of $\vect$. To overcome this difficulty one has to 
glue together various local solutions of the form \eqref{WKBloes}. 
Here the crucial observation is that whenever an eigenvalue of 
$\frac{\partial \vecx}{\partial \vecy}$ changes sign (causing 
$\frac{\partial\vecy}{\partial\vecx}$ to become singular) this results in
a phase jump of the wave function by $-\frac{\pi}{2}$. For a closed curve
$\mathcal{C}$ the number of times this happens along $\mathcal{C}$ is 
a topological invariant of $\mathcal{C}$, its Maslov index $\mu$, see
\cite{Mas72,MasFed81}. On a $d$-torus $\T^d$ let us choose a set of basis 
cycles $\{\mathcal{C}_j\}$, $j=1,\hdots,d$, such that along $\mathcal{C}_j$
the angle $\vartheta_j$ increases by $2\pi$. and all other angles 
$\vartheta_{k\neq j}$ remain constant. See figure \ref{fig:torus} for an 
illustration of this basis for a $2$-torus. Then every closed curve on $\T^d$
is a linear combination of the basis cycles $\mathcal{C}_j$.

All we have to do now in order to get a globally well-defined WKB
wave function is to make sure that $\psi_{\rm WKB}(\vecx)$ returns to its 
initial value when we follow its value along $\mathcal{C}_j$. In other 
words the phase change along a loop has to be an integer multiple of $2\pi$,
i.e.
\begin{equation}
\label{gesamt_phase}
  \frac{1}{\hbar} \oint_{\mathcal{C}_j} \vecP \, \ud \vecX
  - \mu_j \frac{\pi}{2} = 2\pi \, n_j \, , \quad n_j \in \Z \, .
\end{equation}
These are the quantisation conditions \eqref{EBK}. Since the action variables
$\vecI$ are given by 
\begin{equation}
\label{action_variables}
  I_j = \frac{1}{2\pi} \oint_{\mathcal{C}_j} \vecP \, \ud\vecX
\end{equation}
the EBK energies read
\begin{equation}
  E_{\vecn} 
  = \overline{H} \left( \hbar \left(\vecn+\frac{\vecmu}{4} \right) \right)
\end{equation}
where $\overline{H}$ is the Hamiltonian transformed to action and angle 
variables and $\mu_j$ denotes the Maslov index of the basis cycle 
$\mathcal{C}_j$. Further restrictions on the values which the integers 
$\vecn$ may assume usually arise from the possible values the action variables 
$\vecI$ can take in the particular problem of classical mechanics, 
cf. the examples given below.

\section{The semiclassical wave function for the Dirac equation}
\label{sec:scDirac}

In this section we review the determination of semiclassical wave 
functions for the Dirac equation and introduce the relevant notation 
for the following sections.

The first steps towards a semiclassical wave function for the Dirac equation 
were undertaken by Pauli \cite{Pau32}. He inserted an ansatz similar to 
\eqref{WKBansatz} into the Dirac equation and found that the phase function
$S$ has to solve a relativistic Hamilton-Jacobi equation. He then solved 
the analogue of the transport equation \eqref{scalar_transport} 
for a particular case 
but did not derive a general expression like \eqref{scalar_a0}. 
The problem was taken up again many years later by Rubinow and Keller 
\cite{RubKel63} who proceeded one step further. They showed that the 
solution of the transport equation can be related to the Thomas 
precession \cite{Tho26,Tho27}, see also \cite{BarMicTel59}, of a classical 
spin vector. However, also in this work no general quantisation conditions 
similar to \eqref{EBK} were given. As we will see in the following sections 
their construction is complicated by the occurrence of non-Abelian Berry 
phases for which additional integrability conditions are needed. 

We also list some related literature which we will, however, not 
directly refer to in the following:
Semiclassical approximations to the radial Dirac equation were studied
in \cite{Goo53,RosYen64,Lu70}. 
The semiclassical time evolution of the Dirac equation was examined in 
\cite{Yaj82a}. Semiclassical quantisation of subspectra of the Dirac 
Hamiltonian based on the complex germ method was discussed in 
\cite{BagBelTriYev94a,BagBelTriYev94b}. The time evolution of semiclassical 
Wigner functions for the Dirac equation is addressed in \cite{Spo00}.

We briefly repeat the basic steps in the derivation of semiclassical wave
functions for the Dirac equation. For details we refer the reader to 
\cite{RubKel63} or to \cite{BolKep99a} where the notation is similar
to that used here.

The aim is to find asymptotic solutions to the stationary Dirac equation 
\begin{equation}
\label{Dirac_eq} 
  \op{H}_\mathrm{D} \Psi(\vecx) = E \Psi(\vecx)
\end{equation}
with Dirac Hamiltonian 
\begin{equation}
\label{Dirac_Hamiltonian}
  \op{H}_\mathrm{D} = 
  c\vecalph \left(\frac{\hbar}{\ui} \nabla - \frac{e}{c} \vecA(\vecx) \right) 
  + \beta mc^2 + e\phi(\vecx)
\end{equation}
in the semiclassical limit $\hbar \to 0$. The wave function is now a 
four-spinor, $\Psi \in L^2(\R^3) \otimes \C^4$, and the $4\times4$ matrices 
$\vecalph$ and $\beta$ are given by 
\begin{equation}
\label{alpha_beta_def}
  \vecalph = \begin{pmatrix} 0 & \vecsig \\ \vecsig & 0 \end{pmatrix}
  \, , \quad 
  \beta =  \begin{pmatrix} \eins_2 & 0 \\ 0 & -\eins_2 \end{pmatrix} 
  \, , 
\end{equation}
where each entry is to be understood as a $2\times2$ matrix. 
The Pauli matrices $\vecsig$ are given by 
\begin{equation}
  \sigma_1 = \begin{pmatrix} 0 & 1 \\ 1 & 0 \end{pmatrix}
  \, , \quad 
  \sigma_2 = \begin{pmatrix} 0 & -\ui \\ \ui & 0 \end{pmatrix}
  \, , \quad 
  \sigma_3 = \begin{pmatrix} 1 & 0 \\ 0 & -1 \end{pmatrix}
\end{equation}
and $\eins_n$ denotes the $n\times n$ unit matrix. The Dirac equation 
describes a particle of mass $m$ and charge $e$ moving under the influence of
the external electro-magnetic potentials $(\phi(\vecx),\vecA(\vecx))$, i.e.
we have fixed a frame of reference in which the potentials are static.

We now modify the semiclassical ansatz \eqref{WKBansatz} such that the 
amplitudes $a_k(\vecx)$ take values in $\C^4$ but the phase $S(\vecx)$
is kept scalar. Inserting this ansatz into the Dirac equation 
\eqref{Dirac_eq} yields
\begin{equation}
\label{insert_scansatz}
\begin{split}
  \left[ H_\mathrm{D}(\nab{x}S,\vecx) - E \right] a_0 
  + \frac{\hbar}{\ui} \big\{ 
    \left[ H_\mathrm{D}(\nab{x}S,\vecx) - E \right] a_1
    + c\vecalph (\nab{x}a_0) 
  \big\} + \O(\hbar^2) = 0 \, ,
\end{split}
\end{equation}
where $H_\mathrm{D}(\vecp,\vecx)$ denotes the Weyl symbol of the 
Dirac Hamiltonian 
\eqref{Dirac_Hamiltonian}, 
\begin{equation}
  H_\mathrm{D}(\vecp,\vecx) = \begin{pmatrix} 
  mc^2 + e\phi(\vecx) & 
  \vecsig \left(c\vecp - e\vecA(\vecx) \right) \\
  \vecsig \left(c\vecp - e\vecA(\vecx) \right) &
  mc^2 - e\phi(\vecx) \end{pmatrix} .
\end{equation}
When comparing \eqref{insert_scansatz} with \eqref{insertWKBansatz} notice 
that $c\vecalph = \nab{p}H_\mathrm{D}$. Since the leading order equation 
\begin{equation}
\label{leading_order}
  \left[ H_\mathrm{D}(\nab{x}S,\vecx) - E \right] a_0 = 0 
\end{equation}
is a matrix equation it implies the necessary condition that the expression 
in square brackets has an eigenvalue which vanishes identically, i.e.
\begin{equation}
\label{HJG+-}
  H^\pm(\nab{x}S,\vecx) = E \, , 
\end{equation}
where the eigenvalues
\begin{equation}
\label{H+-}
  H^\pm(\vecp,\vecx) 
  = e\phi(\vecx) \pm \sqrt{(c\vecp-e\vecA(\vecx))^2 + m^2c^4}
\end{equation}
of $H_\mathrm{D}$ act as classical Hamiltonians for our problem. 
In the following we will use the notation $S^\pm$ indicating which of 
the two Hamilton-Jacobi equations \eqref{HJG+-} is solved by the 
respective phase function; the solution itself is again given by 
\eqref{S_integrated} where now the integration is along the flow lines of 
$\phi_{H^\pm}^t$.
Each of the eigenvalues \eqref{H+-} has a multiplicity of two and the 
corresponding eigenvectors can be chosen as the columns of the 
$4\times2$ matrices $V_\pm$, 
\begin{equation}
\begin{split} 
  V_+(\vecp,\vecx) 
  &= \frac{1}{\sqrt{2\varepsilon(\varepsilon+mc^2)}}
  \begin{pmatrix} \varepsilon + mc^2 \\ \vecsig (c\vecp-e\vecA(\vecx)) 
  \end{pmatrix} , \\
  V_-(\vecp,\vecx)
  &= \frac{1}{\sqrt{2\varepsilon(\varepsilon+mc^2)}}
  \begin{pmatrix} \vecsig (c\vecp-e\vecA(\vecx)) \\ -(\varepsilon + mc^2)
  \end{pmatrix} ,
\end{split}
\end{equation}
where we have introduced the abbreviation 
\begin{equation}
\label{epsilon_def}
  \varepsilon(\vecp,\vecx) := \sqrt{(c\vecp-e\vecA(\vecx))^2 + m^2c^4} \, .
\end{equation}
With this choice the eigenvectors are orthonormal and complete, i.e. 
\begin{equation}
\begin{split} 
   V_\pm^\dag(\vecp,\vecx) \, V_\pm(\vecp,\vecx) = \eins_2 \, , \quad
   V_\mp^\dag(\vecp,\vecx) \, V_\pm(\vecp,\vecx) = 0 \, , \\
   \text{and} \quad 
   V_+(\vecp,\vecx) V_+^\dag(\vecp,\vecx) 
   + V_-(\vecp,\vecx) V_-^\dag(\vecp,\vecx) = \eins_4 \, .
\end{split}
\end{equation} 
Now $S^\pm(\vecx)$ solving one of the Hamilton-Jacobi equations \eqref{HJG+-} 
is not sufficient in order to satisfy eq.~\eqref{leading_order}. 
Instead we also need $a_0^\pm(\vecx)$ to be of the form 
\begin{equation}
\label{def_b}
  a_0^\pm(\vecx) = V_\pm(\nab{x}S,\vecx) \, b^\pm(\vecx)
\end{equation}
with the still unknown function $b^\pm(\vecx)$ taking values in $\C^2$.

An equation for $b^\pm$ can be obtained from the next-to-leading order
equation deriving from \eqref{insert_scansatz} by inserting \eqref{def_b} 
and multiplying with $V_\pm^\dag(\nab{x}S,\vecx)$ from the left yielding
\begin{equation}
  c V_\pm^\dag(\nab{x}S,\vecx) \vecalph V(\nab{x}S,\vecx) \, (\nab{x} b^\pm)
  + c V_\pm^\dag(\nab{x}S,\vecx) \vecalph [(\nab{x}V)(\nab{x}S,\vecx)] \, b^\pm
  = 0 \, .
\end{equation}
After some algebra one arrives at 
\begin{equation}
\label{transport_b}
\begin{split} 
  &\left[ (\nab{p}H^\pm)(\nab{x}S,\vecx) \right](\nab{x} b^\pm) 
  + \frac{1}{2} \left[ \nab{x} (\nab{p}H^\pm)(\nab{x}S^\pm,\vecx) \right] 
    b^\pm 
  + \frac{\ui}{2} \vecsig \vcB^\pm(\nab{x}S^\pm,\vecx) \, b^\pm
  = 0 \\
  &\ \text{with} \quad \vcB^\pm(\vecp,\vecx) 
  = \mp \frac{ec}{\varepsilon} \vecB(\vecx) 
    + \frac{ec}{\varepsilon(\varepsilon+mc^2)} \,  
      [c\vecp-e\vecA(\vecx)] \times \vecE(\vecx) \, .
\end{split} 
\end{equation}
Here we have introduced the electric and magnetic fields 
\begin{equation}
  \vecE(\vecx) = -\nabla\phi(\vecx) 
  \quad \text{and} \quad 
  \vecB(\vecx) = \nabla \times \vecA(\vecx) \, .
\end{equation}
Equation \eqref{transport_b} is of the same form as the scalar transport 
equation \eqref{scalar_transport} except for the extra term 
$\frac{\ui}{2} \vecsig\vcB^\pm \, b^\pm$. Thus we can exploit our 
knowledge of how to solve \eqref{scalar_transport} by making the ansatz 
\begin{equation}
  b^\pm(\vecx) = \sqrt{\det\frac{\partial\vecy}{\partial\vecx}} \, u^\pm(\vecx)
\end{equation}
leaving us with the spin transport equation 
\begin{equation}
  \dot{u}^\pm 
  + \frac{\ui}{2} \vecsig \vcB^\pm(\nab{x}S^\pm,\vecx) \, u^\pm = 0
\end{equation}
for the $\C^2$-valued function $u^\pm(\vecx)$, 
where the dot denotes a derivative along the Hamiltonian flow 
$\phi_{H^\pm}^t$, cf. eq. \eqref{def_dot}.

Given an initial value $u^\pm(\vecy)$ at a point $\vecy$ its value at a point 
$\vecx$, which is connected to $\vecy$ by the trajectory 
$\phi_{H^\pm}^t(\vecxi,\vecy)$, is given by 
\begin{equation}
  u^\pm(\vecx) = d_\pm(\vecxi,\vecy,t) \, u^\pm(\vecy) \, , 
\end{equation}
where $d_\pm(\vecxi,\vecy,t)$ is a $2\times2$ matrix. We have explicitly
indicated the dependence of $d_\pm$ on the initial point in phase space 
$(\vecxi,\vecy)$ where we start the integration and the time $t$ until which
we proceed. Clearly, $d_\pm$ also has to solve a spin transport equation,
\begin{equation}
\label{spin_transport_d}
  \dot{d}_\pm(\vecxi,\vecy,t) 
  + \frac{\ui}{2} \vecsig \vcB^\pm(\phi_{H^\pm}^t(\vecxi,\vecy)) \, 
    d_\pm(\vecxi,\vecy,t) = 0 
  \, , \quad 
  d_\pm(\vecxi,\vecy,0) = \eins_2 \, .
\end{equation}
Since the coefficient $\frac{\ui}{2} \vecsig \vcB^\pm$ takes values in the
Lie algebra $\su(2)$ it follows that $d_\pm(\vecxi,\vecy,0) \in \SU(2)$.
More precisely, the matrix valued function $d_\pm$ is an $\SU(2)$-valued 
cocycle of the flow $\phi_H^t$ as one easily verifies the composition law
\begin{equation}
  d_\pm(\vecxi,\vecy,t+t^\prime) 
  = d_\pm(\phi_{H^\pm}^t(\vecxi,\vecy),t^\prime) \, d_\pm(\vecxi,\vecy,t) \, .
\end{equation}
Accordingly we can define the skew product flow \cite{BolKep99b}
\begin{equation}
\label{skew_def}
\begin{split} 
  Y_\pm^t: \
  \R^d \times \R^d \times \SU(2) \ &\to \ \R^d \times \R^d \times \SU(2)\\
  (\vecp,\vecx,g) \quad &\mapsto 
    \left( \phi_{H^\pm}^t(\vecp,\vecx), d_\pm(\vecp,\vecx,t)g \right)
\end{split}
\end{equation}
which preserves the product of Liouville measure on phase space 
$\R^d\times\R^d$ and Haar measure on $\SU(2)$, see \cite{BolKep99b}.
Since the cocycle $d_\pm$ takes values in the group $G=\SU(2)$ this 
construction is also known as a group extension or, more precisely
as an $\SU(2)$-extension.

At this point the spin degree of freedom is still described on a quantum 
mechanical level in the sense that it is represented by elements of $\C^2$
which are evolved in time by an $\SU(2)$-valued propagator. It is, however,
possible to switch to a purely classical description. To this end 
consider the adjoint representation of $\SU(2)$ defined by 
\begin{equation}
\label{adjoint_rep}
\begin{split}
  \Ad_g : \su(2) &\to \su(2)\\
  Z & \to gZg^{-1} \, .
\end{split}
\end{equation}
Since an element of the Lie algebra $\su(2)$ can be written as a linear
combination of the Pauli matrices, $Z=\vecz\vecsig$, $\vecz\in\R^3$, 
expanding also the right hand side of \eqref{adjoint_rep} in this basis,
$gZg^{-1}=\big(\varphi(g)\vecz\big)\vecsig$, provides us with a rotation
matrix $\varphi(g)$. The map $\varphi:\SU(2)\to\SO(3)$ is two-to-one and 
known as the covering map. Anticipating classical spin as a vector of 
constant length let us consider $\vecs_0 \in S^2 \hookrightarrow \R^3$.
On easily verifies that $\vecs=\varphi(d(\vecxi,\vecy,t)) \, \vecs_0$
solves 
\begin{equation}
\label{Thomas_prec}
 \dot{\vecs} = \vcB^\pm(\phi_{H^\pm}^t(\vecxi,\vecy)) \times \vecs
\end{equation}
i.e. the equation of Thomas precession \cite{Tho26,Tho27}, which has thus been 
derived from the Dirac equation \cite{RubKel63,BolKep98,BolKep99a,Spo00}.
Classical spin precession \eqref{Thomas_prec} and the Hamiltonian flow
$\phi_{H^\pm}^t$ can now be combined into the classical skew product
\cite{BolGlaKep01} 
\begin{equation}
\label{skew_cl_def}
\begin{split} 
  Y_{\rm cl\pm}^t: 
  \R^d \times \R^d \times S^2\ &\to \ \R^d \times \R^d \times S^2\\
  (\vecp,\vecx,\vecs) \quad &\mapsto 
    \left( \phi_{H^\pm}^t(\vecp,\vecx),
           \varphi(d_\pm(\vecp,\vecx,t)) \vecs \right)  \, .
\end{split}
\end{equation}
As opposed to $Y_{\pm}^t$ this is a symplectic flow conserving the 
product of Liouville measure on the phase space $\R^d\times\R^d$ of
the translational degrees of freedom and 
the surface element on the sphere $S^2$ which acts 
as the phase space for the spin degree of freedom. As we will see below 
the properties of $Y_{\rm cl\pm}^t$ determine the semiclassical solutions
of the Dirac equation as the properties of $\phi_{H^\pm}^t$ determine
that of the Schr\"odinger equation.

Collecting the results of the previous paragraphs, so far we have obtained
the following expression for the semiclassical wave function for the Dirac 
equation,
\begin{equation}
\label{sc_wave_function_Dirac}
  \Psi_{\rm sc}^\pm(\vecx) \sim 
  \sqrt{\det\frac{\partial\vecy}{\partial\vecx}} \, 
  \ue^{\frac{\ui}{\hbar}S^\pm(\vecx)} \,
  V_\pm(\nab{x}S^\pm,\vecx) \, d_\pm(\vecxi,\vecy,t) \, u^\pm(\vecy) \, .
\end{equation}
The new ingredients as compared to \eqref{WKBloes} are the projection 
matrices $V_\pm$ selecting either positive or negative kinetic energies
and the spinor part $d_\pm(\vecxi,\vecy,t) \, u^\pm(\vecy)$. We can now
explain why \eqref{sc_wave_function_Dirac} does not immediately lead
to a generalisation of the EBK quantisation conditions \eqref{gesamt_phase}.
Assume that the Hamiltonian flow generated by either $H^+(\vecp,\vecx)$ 
or $H^-(\vecp,\vecx)$ is integrable in the sense of Theorem 
\ref{Liouville-Arnold}. Then we know that the factors 
$\sqrt{\det\frac{\partial\vecy}{\partial\vecx}}$ and 
$\ue^{\frac{\ui}{\hbar}S^\pm(\vecx)}$ can be defined globally by the 
construction described in the preceeding section. This also applies to
the projection matrix $V_\pm$ which simply has to be evaluated along the 
respective path of integration; in particular, for a closed path also the
value of $V_\pm$ returns to its initial value thus not contributing to the 
quantisation conditions.

In contrast, integrating the spin transport equation \eqref{spin_transport_d}
along a closed path $\mathcal{C}_j$ will yield an $\SU(2)$ matrix $d_j$.
Thus the initial and final values of the semiclassical wave function
\eqref{sc_wave_function_Dirac} would not just differ by a phase factor 
but by an $\SU(2)$ transformation which would have to be compensated for in 
order to turn $\Psi_{\rm sc}(\vecx)$ into a globally single-valued
object. Even worse, integration along a different loop $\mathcal{C}_{k\neq j}$
(which we still have to define, so far we are only able to integrate 
\eqref{spin_transport_d} along the flow lines of $\phi_{H^\pm}^t$) will in 
general yield a different $\SU(2)$ matrix $d_{k\neq j}$ which need not 
commute with the first one. This can make it impossible to find a globally 
well-defined semiclassical solution as was pointed out by Emmrich and 
Weinstein in a general setting \cite{EmmWei96}. From the point of view 
of physics this is not surprising. We know that in order to be able 
to use EBK quantisation the corresponding classical system has to be
integrable. Now we are dealing with a system that has an additional 
degree of freedom, namely spin. So far, however, we have not imposed any
condition on the spin dynamics but we have only required the Hamiltonian flow
$\phi_{H^\pm}^t$ to be integrable. In the following section we will show how 
the notion of integrability can be extended to skew products of the form
\eqref{skew_def} or \eqref{skew_cl_def} and prove a generalisation of 
Theorem \ref{Liouville-Arnold}.

\section{Integrability of skew products}
\label{sec:integrable_skew}

From the point of view of semiclassics the crucial aspect in the 
characterisation of integrable systems by Theorem \ref{Liouville-Arnold} 
is the geometrical description of the invariant manifolds which is a 
consequence of the existence of commuting flows $\phi_1^t,\hdots,\phi_d^t$.
Therefore, one has to find a generalisation of the condition 
\eqref{involution}. We will show that in this sense the following definition
provides a good generalisation of the notion of integrability. 
\begin{definition}
\label{def_integrable_Ycl}
The skew product $Y_{\rm cl\pm}^t$ is called integrable, if 
\begin{enumerate}
\item[\rm (i)] the underlying Hamiltonian flow $\phi_{H^\pm}^t$ is 
           integrable in the sense of Liouville and Arnold 
           (Theorem \ref{Liouville-Arnold}), i.e. besides the Hamiltonian 
           $H^\pm(\vecp,\vecx)=:A_1(\vecp,\vecx)$ there are $d-1$ more 
           independent integrals of motion, 
           $A_2(\vecp,\vecx),\hdots,A_d(\vecp,\vecx)$ with 
           \begin{equation}
             \{A_j,A_k\} = 0 \quad \forall \ j,k=1,\hdots, d
           \end{equation}
           and 
\item[(ii)] the flows $\phi_2^t,\hdots,\phi_d^t$ 
            can also be extended to skew products $\Ycl_j^t$ on 
            $\R^d \times \R^d \times S^2$ ($Y_{\rm cl \pm}^t \equiv \Ycl_1^t$)
            with fields $\vcB_j(\vecp,\vecx)$, i.e
            \begin{equation}
            \label{other_skew}
              \Ycl_j^t(\vecp,\vecx,\vecs) 
              = \left( \phi_j^t(\vecp,\vecx), 
                \varphi(d_j(\vecp,\vecx,t)) \vecs \right)
            \end{equation}
            \begin{equation} 
              \dot{d}_j(\vecp,\vecx,t) 
              + \frac{\ui}{2}\vecsig\vcB_j(\phi_j^t(\vecp,\vecx)) \, 
                d_j(\vecp,\vecx,t) = 0 
              \, , \quad d_j(\vecp,\vecx,0) = \eins_2 \, ,   
            \end{equation} 
            fulfilling 
            \begin{equation}
            \label{spin_involution}
            \{A_j,\vcB_k\} + \{\vcB_j,A_k\} - \vcB_j \times \vcB_k = 0 
            \quad \forall \ j,k = 1,\hdots,d  \, . 
            \end{equation}
\end{enumerate}
\end{definition}
\noindent
In view of what was said above let us first show that this is indeed a good
definition by the following lemma.
\begin{lemma}
\label{commuting_skew}
Two skew products $\Ycl_j^t$ and $\Ycl_k^t$ of the type \eqref{other_skew} 
commute if and only if the corresponding base flows $\phi_j^t$ and $\phi_k^t$
commute and if the fields $\vcB_j$ and $\vcB_k$ fulfil 
\begin{equation}
\label{spin_involution_repeat}
  \{ A_j, \vcB_k \} + \{ \vcB_j, A_k \} - \vcB_j \times \vcB_k = 0 \, .
\end{equation}
\end{lemma}
\noindent
\proof
Since the two skew products can only commute if the corresponding base
flows commute it remains to show that 
\begin{equation}
  \varphi\big(d_j(\phi_k^t(\vecp,\vecx),t^\prime)\big) \,
  \varphi(d_k(\vecp,\vecx,t)) \, \vecs
  = \varphi\big(d_k(\phi_j^{t^\prime}(\vecp,\vecx),t)\big) \,
  \varphi(d_j(\vecp,\vecx,t^\prime)) \, \vecs
\end{equation}
for all $(\vecp,\vecx,\vecs) \in \R^d \times \R^d \times S^2$ and all 
$t,t^\prime \in \R$, or equivalently
\begin{equation}
  \varphi\big(d_j(\phi_k^t(\vecp,\vecx),t^\prime)\big) \,
  \varphi(d_k(\vecp,\vecx,t))
  = \varphi\big(d_k(\phi_j^{t^\prime}(\vecp,\vecx),t)\big) \,
  \varphi(d_j(\vecp,\vecx,t^\prime))
\end{equation}
for all $(\vecp,\vecx) \in \R^d \times \R^d$ and all $t,t^\prime \in \R$.
Moreover, since $\varphi$ is a double covering, \eqref{spin_involution_repeat}
is also equivalent to 
\begin{equation}
  d_j(\phi_k^t(\vecp,\vecx),t^\prime) \,
  d_k(\vecp,\vecx,t) \, 
  \left[d_k(\phi_j^{t^\prime}(\vecp,\vecx),t) \,
        d_j(\vecp,\vecx,t^\prime) \right]^{-1}
  = \pm \eins_2 \, .
\end{equation}
However, due to 
\begin{equation}
  \left. 
  d_j(\phi_k^t(\vecp,\vecx),t^\prime) \,
  d_k(\vecp,\vecx,t) \, 
  \left[d_k(\phi_j^{t^\prime}(\vecp,\vecx),t) \,
        d_j(\vecp,\vecx,t^\prime) \right]^{-1} \right|_{t^\prime=0}
  = + \eins_2 \, .
\end{equation}
we conclude that
\begin{equation}
  \eqref{spin_involution_repeat} \quad \Leftrightarrow \quad
  \Delta(t,t^\prime) = 0 \, , 
\end{equation}
where the difference $\Delta(t,t^\prime)$ is defined by
\begin{equation}
\label{Delta_def}
  \Delta(t,t^\prime) :=
  d_j(\phi_k^t(\vecp,\vecx),t^\prime) \,
  d_k(\vecp,\vecx,t) - 
  d_k(\phi_j^{t^\prime}(\vecp,\vecx),t) \,
  d_j(\vecp,\vecx,t^\prime) \, .
\end{equation}
It is easy to see that $\Delta(t,0) = \Delta(0,t^\prime) = 0$ 
$\forall \ t,t^\prime$ and thus 
\begin{equation}
\label{Delta_um_Null}
  \Delta(t,t^\prime) = \frac{\partial^2\Delta}{\partial t \partial t^\prime}
  (0,0) \, tt^\prime + \o(t^2+t^{\prime2})
  \, , \quad t,t^\prime \to 0 \, .
\end{equation}
The relevant second derivative is given by 
\begin{equation}
\label{Delta''}
\begin{split}   
  \frac{\partial^2\Delta}{\partial t\partial t^\prime}(0,0)
  &= \frac{\partial}{\partial t} \left[ 
    -\frac{\ui}{2} \vecsig \vcB_j(\phi_k^t(\vecp,\vecx)) \, d_k(\vecp,\vecx,t) 
    \right]_{t=0}\\
  &\quad
    - \frac{\partial}{\partial t^\prime} \left[ 
    -\frac{\ui}{2} \vecsig \vcB_k(\phi_j^{t^\prime}(\vecp,\vecx)) \, 
        d_j(\vecp,\vecx,t^\prime) \right]_{t^\prime=0}\\
  &= \left[ - \frac{\ui}{2} \vecsig \{ A_k, \vcB_j \}(\vecp,\vecx) 
        + \frac{\ui}{2} \big( \vecsig \vcB_j(\vecp,\vecx) \big) \, 
          \frac{\ui}{2} \big( \vecsig \vcB_k(\vecp,\vecx) \big) \right]\\
  &\quad 
     - \left[ - \frac{\ui}{2} \vecsig \{ A_j, \vcB_k \}(\vecp,\vecx) 
        + \frac{\ui}{2} \big( \vecsig \vcB_k(\vecp,\vecx) \big) \, 
          \frac{\ui}{2} \big( \vecsig \vcB_j(\vecp,\vecx) \big) \right]\\
  &= \frac{\ui}{2} \vecsig \big[
     \{ \vcB_j, A_k \} + \{ A_j, \vcB_k \} - \vcB_j \times \vcB_k 
     \big](\vecp,\vecx) \, , 
\end{split}     
\end{equation}
which already proves one half of Lemma \ref{commuting_skew}: 
If the skew products $\Ycl_j^t$ and $\Ycl_k^t$ commute then
\eqref{spin_involution_repeat} holds. For the reverse direction notice that 
\eqref{spin_involution_repeat} now implies 
\begin{equation}
  \Delta(t,t^\prime) = \o(t^2+t^{\prime2}) \, , \quad t,t^\prime \to 0 \, .
\end{equation}
Dividing the time intervals $t$ and $t^\prime$ into $N$ subintervals of
length $\varepsilon=t/N$ and $\varepsilon^\prime=t^\prime/N$, respectively,
we can rewrite the first term in \eqref{Delta_def} as follows, 
\begin{equation}
\begin{split} 
  d_j(\phi_k^t&(\vecp,\vecx),t^\prime) \, d_k(\vecp,\vecx,t)\\
  &= d_j(\phi_k^t\circ\phi_j^{\varepsilon^\prime}(\vecp,\vecx),
        t^\prime-\varepsilon^\prime) \, 
    \underbrace{
    d_j(\phi_k^t(\vecp,\vecx),\varepsilon^\prime) \, 
    d_k(\phi_k^{t-\varepsilon}(\vecp,\vecx),\varepsilon)}
    _{=:(*)}
    \, d_k(\vecp,\vecx,t-\varepsilon)\\
  (*)
  &=d_k(\phi_k^{t-\varepsilon}\circ\phi_j^{\varepsilon^\prime}(\vecp,\vecx),
        \varepsilon) \, 
    d_j(\phi_k^{t-\varepsilon}(\vecp,\vecx),\varepsilon^\prime)
    + \o(\varepsilon^2+\varepsilon^{\prime2}) \, .
\end{split}
\end{equation}
Repeating this procedure $N^2$ times yields 
\begin{equation}
\label{Delta_vanishes}
  \Delta(t,t^\prime) 
  = N^2 \, \o(\varepsilon^2+\varepsilon^{\prime2})
  = N^2 \, \o(1/N^2) = \o(1) 
  \, , \quad N\to\infty \, ,
\end{equation}
i.e. $\Delta(t,t^\prime)$ vanishes which proves Lemma \ref{commuting_skew}.
\hfill $\Box$

Having thus established a reasonable generalisation of the notion of
integrability we can now state the main result of this section.
\begin{theorem}
\label{integrable_Ycl}
If the skew product flow $Y_{\rm cl\pm}^t$ is integrable,
the combined phase space $\R^d \times \R^d \times S^2$ can be
decomposed into invariant bundles 
$\mathcal{T}_\theta \stackrel{\pi}{\longrightarrow} \T^d$ over 
Liouville-Arnold tori $\T^d$ with typical fibre $S^1$. 
The bundles can be embedded
in $\T^d \times S^2$ such that the fibres are characterised by the 
latitude with respect to a local direction $\vecn(\vecp,\vecx)$, i.e.
\begin{equation}
\label{T_theta}
  \mathcal{T}_\theta = \{ (\vecp,\vecx,\vecs) \in \T^d \times S^2 \, | \,
        \sphericalangle (\vecs,\vecn(\vecp,\vecx)) = \theta \} \, . 
\end{equation}
\end{theorem}
\noindent
As we have seen above, integrability of $Y_\mathrm{cl}^t$ also implies
similar properties of $Y^t$. Let us therefore state the following proposition
for group extensions, the proof of which will facilitate the proof of Theorem
\ref{integrable_Ycl}.
\begin{proposition}
\label{Gext_int}
Let $\phi_1^t$ be an integrable Hamiltonian flow in the sense of Theorem 
\ref{Liouville-Arnold}. A group extension $Y_1^t$ of $\phi_1^t$ with group
$G=\U(n)$ or a subgroup thereof, 
\begin{equation}
\begin{split}
  Y_1^t: \ \R^d\times\R^d\times G \ &\to \quad \R^d\times\R^d\times G \\
         (\vecp,\vecx,g) \quad &\mapsto \
         \left( \phi_1^t(\vecp,\vecx),d(\vecp,\vecx,t) \, g \right) \, ,
\end{split}
\end{equation}
\begin{equation}
\label{transport_allgemein}
  \dot{d}(\vecp,\vecx,t) 
  + M\!\left(\phi_1^t(\vecp,\vecx)\right) \, d(\vecp,\vecx,t) = 0 \, , \quad 
  d(\vecp,\vecx,0) = \eins_n \, , \quad 
  M: \ \R^d\times\R^d \to \mathfrak{g} \, , 
\end{equation} 
where $\mathfrak{g}$ is the Lie-albegra of $G$, 
is called integrable if the flows $\phi_1^t,\hdots,\phi_d^t$ can also be 
extended to $G$-extensions $Y_2^t,\hdots,Y_d^t$ with 
\begin{equation}
\label{M_involution}
\{A_j,M_k\} + \{M_j,A_k\} + [M_j,M_k] = 0 
\quad \forall \ j,k = 1,\hdots,d  \, . 
\end{equation}
Then \eqref{transport_allgemein} defines a connection in the trivial principal
bundle $\T^d \times G$ whose holonomy group is an Abelian subgroup of $G$.
\end{proposition}
\noindent
We remark that in the case of $G=\SU(2)$ the matrices $M_j$ take the form 
\begin{equation}
  M_j(\vecp,\vecx) = \frac{\ui}{2}\vecsig\vcB_j(\vecp,\vecx)
\end{equation}
thus reducing condition \eqref{M_involution} to \eqref{spin_involution}.\\
\proofof{Proposition \ref{Gext_int}}
First notice that condition \eqref{M_involution} ensures the commutativity
of the $G$-extensions $Y_1^t,\hdots,Y_d^t$. This can be seen by repeating
the proof of Lemma \ref{commuting_skew} where in the definition 
\eqref{Delta_def} of the difference $\Delta(t,t^\prime)$ the matrices 
$d_j$ and $d_k$ take values in $G$, see \eqref{transport_allgemein}.
Then equation \eqref{Delta_um_Null} is still valid and \eqref{Delta''}
reads
\begin{equation}
\begin{split}   
  \frac{\partial^2\Delta}{\partial t\partial t^\prime}(0,0)
  &= \big( \{ M_j, A_k \} + \{ A_j, M_k \} + [M_j,M_k] \big) (\vecp,\vecx) 
  \, .
\end{split}     
\end{equation}
The following steps up to \eqref{Delta_vanishes} can be adopted identically.

Now define the multi-time flow
\begin{equation}
  \multiY^\vect := Y_d^{t_d} \circ \cdots \circ Y_1^{t_1} \, , 
\end{equation}
where due to commutativity of the flows $Y_1^{t_1},\hdots,Y_d^{t_d}$ ordering
is irrelevant. Explicitly we have
\begin{align} 
  \multiY^\vect(\vecp,\vecx,g) &= 
  \left( (\phi_d^{t_d} \circ \cdots \circ \phi_1^{t_1})(\vecp,\vecx), \, 
  \multid(\vecp,\vecx,\vect) \, g \right) \quad \text{with}\\
\label{Def_multid}
  \multid(\vecp,\vecx,\vect):&= 
  d_d\left( (\phi_{d-1}^{t_{d-1}} \circ \cdots \circ \phi_1^{t_1}) 
        (\vecp,\vecx), t_d \right) \cdots
  d_1( \vecp,\vecx, t_1) g \, . 
\end{align}
Consider a Liouville-Arnold torus $\T^d$ which is invariant under the 
restriction of $\multiY^\vect$ to $\R^d\times\R^d$. The multi-time cocycle
\eqref{Def_multid} then defines a connection in $\T^d \times G$. In order to
determine the holonomy of this connection choose a basis 
$\{\mathcal{C}_j\}$ of closed loops on $\T^d$ as described in section 
\ref{sec_ebk}. With each loop $\mathcal{C}_j$ we can associate a unique 
(minimal) tuple $\vect_j$ such that 
\begin{equation}
  \Phi^\vect(\vecp,\vecx) \, , \quad
  t_k \in [0,(t_j)_k] \, , \quad k=1,\hdots,d 
\end{equation}
topologically describes $\mathcal{C}_j$ for any $(\vecp,\vecx) \in \T^d$. 
We denote the cocycle $\multid(\vecp,\vecx,\vect_j)$ 
associated with $\mathcal{C}_j$ by $d_j(\vecp,\vecx)$, i.e.
\begin{equation}
  \multiY^{\vect_j}(\vecp,\vecx,g) = (\vecp,\vecx,d_j(\vecp,\vecx) \, g) \, .
\end{equation}
We immediately see that due to the commutativity of the skew products 
$Y_j^t$ the cocycles $d_j(\vecp,\vecx)$ commute thus generating the Abelian 
subgroup 
\begin{equation}
\label{holonomy_group}
  H_{(\vecp,\vecx)} = \left\{ 
  g  \in G \, \left| \, 
  g=\prod_{j=1}^d [d_j(\vecp,\vecx)]^{n_j} \, ,
  \vecn \in \Z^d \right. \right\}
\end{equation}
of $G$. In order to see how two such groups, say $H_{(\vecp,\vecx)}$ and 
$H_{(\vecp^\prime,\vecx^\prime)}$, are related, recall that for any two points 
$(\vecp,\vecx)$ and $(\vecp^\prime,\vecx^\prime)$ on a Liouville-Arnold torus 
there exists a $d$-tuple $\vect$ such that 
\begin{equation}
  \Phi^\vect(\vecp,\vecx) = (\vecp^\prime,\vecx^\prime) \, .
\end{equation}
Again due to commutativity we have the equality 
\begin{equation}
  \multiY^{\vect_j} 
  = \multiY^{-\vect} \circ \multiY^{\vect_j}  \circ \multiY^{\vect}
\end{equation}
implying 
\begin{equation}
\label{d_j_konjugiert}
  d_j(\vecp,\vecx) 
  = \multid(\vecp^\prime,\vecx^\prime,-\vect) \, 
    d_j(\vecp^\prime,\vecx^\prime) \, \multid(\vecp^\prime,\vecx^\prime,\vect)
  \, .
\end{equation}
Moreover, 
\begin{equation}
  \multiY^{-\vect} \circ \multiY^{\vect} = \id 
  \quad \Rightarrow \quad
  \multid(\vecp^\prime,\vecx^\prime,-\vect) 
  = [  \multid(\vecp^\prime,\vecx^\prime,\vect) ]^{-1}
\end{equation}
and thus 
\begin{equation}
  H_{(\vecp^\prime,\vecx^\prime)} = g H_{(\vecp,\vecx)} g^{-1}
  \, , \quad 
  g:=\multid(\vecp^\prime,\vecx^\prime,\vect) \, , 
\end{equation}
i.e. the subgroups $H_{(\vecp^\prime,\vecx^\prime)}$ at different points are 
obtained by conjugation with a group element $g \in G$. Thus 
$H_{(\vecp,\vecx)}$ and $H_{(\vecp^\prime,\vecx^\prime)}$ are isomorphic 
and we have identified the Abelian group \eqref{holonomy_group} 
as the holonomy group of the connection defined by \eqref{transport_allgemein}.
\hfill $\Box$

\begin{figure}[t]
\begin{center}
\includegraphics[width=15cm]{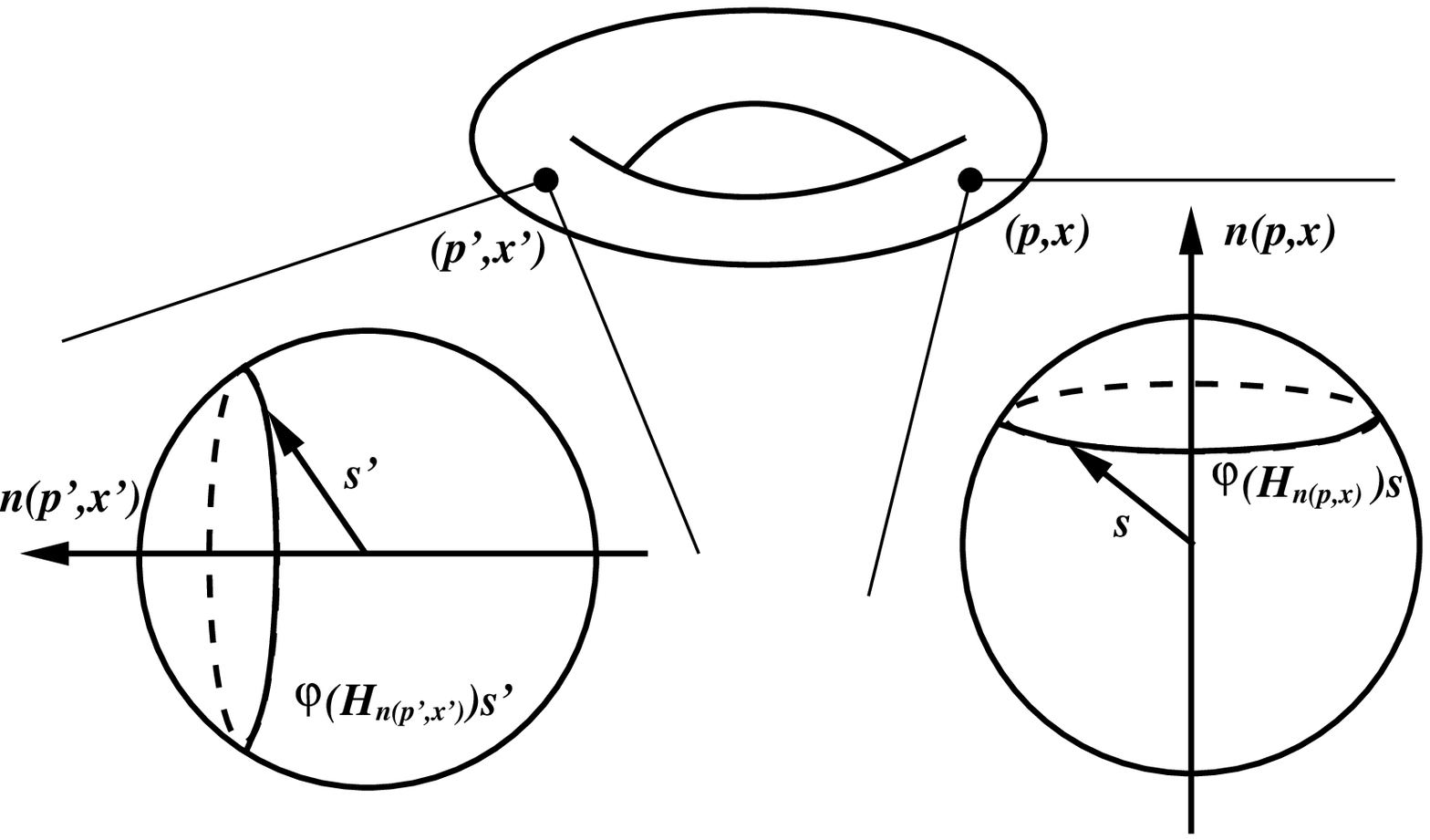}
\caption{Illustration of the invariant manifolds $\mathcal{T}_\theta$, see
\eqref{T_theta}. At different points of the Liouville-Arnold torus 
$\T^d$ spin vectors are restricted to parallels of latitude with different 
axes $\vecn$; the angle 
$\sphericalangle(\vecs,\vecn(\vecp,\vecx)) 
 =  \sphericalangle(\vecs^\prime,\vecn(\vecp^\prime,\vecx^\prime))$
is conserved}
\label{fig:torus_spheres}
\end{center}
\end{figure}

\noindent
\proofof{Theorem \ref{integrable_Ycl}}
Applying Proposition \ref{Gext_int} to the $\SU(2)$-extension \eqref{skew_def},
with each point $(\vecp,\vecx)$ on a Liouville-Arnold torus is 
associated an Abelian subgroup $H_{(\vecp,\vecx)}$ of $\SU(2)$. 
Abelian subgroups of $\SU(2)$ are either one-parameter subgroups or 
discrete subgroups thereof, i.e. we can associate with each point 
$(\vecp,\vecx)\in\T^d$ a one-parameter subgroup of $\SU(2)$. The latter 
can be parametrised as 
\begin{equation}
  H_\vecn := \left\{ 
  g \in \SU(2) \, \left| \, 
  g=\ue^{-\ui\frac{\alpha}{2}\vecsig\vecn} \, , 
  \alpha \in [0,4\pi) \right. \right\}
\end{equation}
with a direction characterised by the unit vector 
$\vecn \in S^2 \hookrightarrow \R^3$. By means of the covering map $\varphi$, 
see definition below \eqref{adjoint_rep}, this construction uniquely determines
a one-parameter subgroup of $\SO(3)$, $\varphi(H_{\vecn(\vecp,\vecx)})$,
at each point $(\vecp,\vecx)$ of a Liouville-Arnold torus. This fact in turn 
allows for a construction of invariant manifolds of $Y_\mathrm{cl}^t$.
Consider a point $(\vecp,\vecx) \in \T^d$ and a spin vector 
$\vecs \in S^2 \hookrightarrow \R^3$. Transporting $\vecs$ along a path on 
$\T^d$ by means of the multi-time flow 
\begin{equation}
  \multiY_\mathrm{cl}^\vect := \Ycl_d^{t_d} \circ \cdots \circ \Ycl_1^{t_1}:\
  (\vecp,\vecx,\vecs) \ \mapsto \ 
  \left( (\phi_d^{t_d} \circ \cdots \circ \phi_1^{t_1}) (\vecp,\vecx), \, 
         \varphi(\multid(\vecp,\vecx,\vect)) \, \vecs \right)
\end{equation}
gives rise to rotations $\varphi(\multid(\vecp,\vecx,\vect))$ of $\vecs$.
The rotation associated with a closed path $\mathcal{C}_j$
is given by the rotation matrix 
$\varphi(d_j(\vecp,\vecx)) \in \varphi(H_{\vecn(\vecp,\vecx)}) \subset \SO(3)$.
Thus whenever the path on the Liouville-Arnold  torus is closed the spin vector
$\vecs$ returns to a point on the circle 
$\varphi(H_{\vecn(\vecp,\vecx)}) \vecs$, 
which is a parallel of latitude with respect to the axis $\vecn(\vecp,\vecx)$.
Corresponding circles at different points $(\vecp^\prime,\vecx^\prime)$
are obtained as follows, see figure \ref{fig:torus_spheres}.
The one-parameter subgroups at different points of the torus 
are related by conjugation with $\multid(\vecp,\vecx,\vect)$, see 
\eqref{d_j_konjugiert}. However, from the definition of the covering map
$\varphi$, see below \eqref{adjoint_rep}, it follows that 
\begin{equation}
  H_{\vecn(\vecp^\prime,\vecx^\prime)} 
  = \multid(\vecp,\vecx,\vect) \, H_{\vecn(\vecp,\vecx)} \, 
    [\multid(\vecp,\vecx,\vect)]^{-1} 
  = H_{\varphi(\multid(\vecp,\vecx,\vect)) \vecn} \, .
\end{equation}
On the other hand from
\begin{equation}
  \multiY_\mathrm{cl}^\vect(\vecp,\vecx,\vect)
  =\left( \vecp^\prime,\vecx^\prime, \, 
          \varphi(\multid(\vecp,\vecx,\vect)) \vecs \right)
\end{equation}
we see that by moving from $(\vecp,\vecx)$ to $(\vecp^\prime,\vecx^\prime)$
the spin vector $\vecs$ is rotated in the same way as is the axis 
$\vecn$. Therefore the angle 
\begin{equation}
  \theta := \sphericalangle (\vecs,\vecn(\vecp,\vecx))
\end{equation}
is conserved by $\multiY_\mathrm{cl}^\vect$ and thus by $Y_\mathrm{cl}^t$,
and the bundles $\mathcal{T}_\theta$, see \eqref{T_theta}, are invariant, 
concluding the proof of Theorem \ref{integrable_Ycl}.
\hfill $\Box$

\section{Quantisation and spin rotation angles}
\label{sec:rotation_angels}

With the novel notion of integrability for skew products at hand we can now 
return to the semiclassical wave function \eqref{sc_wave_function_Dirac}
of the Dirac equation. We have already seen that except for the spin 
transporter $d_\pm(\vecxi,\vecy,t)$ all terms in the local expression
\eqref{sc_wave_function_Dirac} can be given a global meaning provided that the 
classical translational dynamics is integrable. In the case of the 
Sch\"odinger equation, this observation led to the EBK quantisation 
conditions in a straightforward way, see section \ref{sec_ebk}. We will now 
show that this is also the case for the Dirac equation if the skew 
product flow $Y_{\mathrm{cl}\pm}^t$ is integrable in the sense of Definition 
\ref{def_integrable_Ycl}.

Theorem \ref{integrable_Ycl} provides us with a method for integrating 
the spin transport equation not only along the flow lines of $\phi_{H^\pm}^t$ 
but also along those of the flows $\phi_2^t,\hdots,\phi_d^t$, which are 
defined in the Theorem \ref{Liouville-Arnold}. The resulting spin transporter 
is defined on the whole Liouville-Arnold torus and given by the multi-time
cocycle $\multid(\vecp,\vecx,\vect)$, see \eqref{Def_multid}, for a point 
$(\vecp,\vecx)$ which is reached from $(\vecxi,\vecy)$ by 
$\Phi^\vect(\vecxi,\vecy) = (\vecp,\vecx)$. For a closed path $\mathcal{C}_j$
the initial and final values of the spin part $u^\pm$ of the semiclassical wave
function  are still related by the $\SU(2)$ transformation $d_j(\vecxi,\vecy)$.
However, provided that the skew product $Y_{\mathrm{cl}\pm}^t$ is integrable
we know that the matrices $d_j(\vecxi,\vecy)$ for different loops 
$\mathcal{C}_j$ commute. Therefore we can choose $u^\pm$ to be a simultaneous 
eigenvector of these $\SU(2)$-matrices which effectively reduces the 
$\SU(2)$-holonomy to a simple phase, i.e. a $\U(1)$-holonomy.

Introducing the local axis $\vecn(\vecxi,\vecy)$, see Theorem 
\ref{integrable_Ycl}, we have 
\begin{equation}
\label{def_alpha_j}
  d_j(\vecxi,\vecy) 
  = \ue^{-\frac{\ui}{2}\alpha_j \vecsig\vecn(\vecxi,\vecy)} \, , 
\end{equation}
where $\alpha_j$ is the angle by which a classical spin vector is rotated about
the axis $\vecn(\vecxi,\vecy)$ when transported along $\mathcal{C}_j$;
note that $\alpha_j$ has to measured modulo $4\pi$. The spin part 
$u^\pm(\vecy)$
of the wave function \eqref{sc_wave_function_Dirac} is chosen to be an 
eigenvector of $\vecsig\vecn(\vecxi,\vecy)$ with eigenvalue either $+1$ or
$-1$. Thus the phase shift resulting from spin transport is given by 
$\ue^{\mp\ui\alpha_j/2}$. Together with the contribution from the classical 
action and the Maslov phase, cf. \eqref{gesamt_phase}, the total phase
shift suffered by the wave function when going through the cycle 
$\mathcal{C}_j$ is 
\begin{equation}
  \frac{1}{\hbar} \oint_{\mathcal{C}_j} \vecP \, \ud\vecX 
  - \frac{\pi}{2} \mu_j \mp \frac{\alpha_j}{2} \, .
\end{equation}
Requiring this to be be an integer multiple of $2\pi$ in order to obtain 
a single valued wave function yields the novel quantisation conditions
\begin{equation}
\label{spin_quant}
  \oint_{\mathcal{C}_j} \vecP \, \ud\vecX
  = 2\pi\hbar \left( n_j + \frac{\mu_j}{4} 
                     + m_s \frac{\alpha_j}{2\pi} \right)
\end{equation}
where we have introduced the spin quantum number $m_s=\pm\frac{1}{2}$. 
Equation \eqref{spin_quant} is our central result replacing the EBK 
quantisation condition \eqref{EBK} in the case of relativistic particles with 
spin $\frac{1}{2}$.  If the classical Hamiltonian $H^\pm(\vecp,\vecx)$ 
expressed in action and angle variables $(\vecI,\vecvt)$, see 
\eqref{action_variables}, is given by $\overline{H}(\vecI)$ then 
the semiclassical eigenvalues resulting from \eqref{spin_quant} read
\begin{equation}
  E_{\vecn,m_s}^\pm = \overline{H}\!\left( \hbar 
  \left( \vecn + \frac{\vecmu}{4} + m_s \frac{\vecalph}{2\pi} \right) 
  \right) \, .
\end{equation}

So far we have only employed the flows $\Ycl_j^t$ resulting from the 
``Hamiltonians'' $A_1=H^\pm,A_2,\hdots,A_d$ complemented with fields 
$\vcB_j(\vecp,\vecx)$. For practical purposes it is helpful to directly make 
use of the flows $\phi_{I_j}^t$ generated by the action variables $\vecI$
in the semiclassical quantisation process. To this end these flows have to be 
extended to skew products on $\R^d \times \R^d \times S^2$ by some suitable 
fields $\vcB_{I_j}$. Transport along a basis cycle $\mathcal{C}_j$ is then 
given by $Y_{I_j}^{2\pi}$ and the relevant $\SU(2)$ transformation
$d_j=\exp(-\ui\alpha_j\vecsig\vecn/2)$ is given by the cocycle of just one flow
instead of the a linear combination of the $d$ flows. Since 
\begin{equation}
  \phi_{H^\pm}^t 
  = \phi_{I_d}^{\omega_d t} \circ \cdots \circ \phi_{I_1}^{\omega_1 t} \, , 
\end{equation}
where $\vecomega$ are the fundamental frequencies, see \eqref{theta_punkt}, 
we have to require that 
\begin{equation}
  Y_{\mathrm{cl}\pm}^t \stackrel{!}{=} 
  Y_{I_d}^{\omega_d t} \circ \cdots \circ Y_{I_1}^{\omega_d 1} \, , 
\end{equation}
yielding the consistency condition 
\begin{equation}
\label{consistent_B}
  \vcB^\pm = \sum_{j=1}^d \omega_j \vcB_{I_j} \, . 
\end{equation}
We will illustrate these remarks when applying the method to explicit examples
in sections \ref{sec:spherical} -- \ref{sec:sommerfeld}.

\section{Non-relativistic limit: The Pauli equation}
\label{sec:pauli}

In this section we show that the semiclassical quantisation scheme developed
in the preceding sections for the Dirac equation carries over to the Pauli
equation. The latter arises as a non-relativistic approximation to the Dirac 
equation but can also be generalised to describe particles with spin 
$s\in\N_0/2$ other than $s=\frac{1}{2}$. We show that also in the case 
with $s\neq\frac{1}{2}$ the semiclassical analysis of the Pauli equation 
gives rise to a skew product on $\R^d \times \R^d \times S^2$ which can 
then be quantised along the same lines as in the case of the Dirac equation.
We keep the presentation short relying heavily on our treatment of the 
Dirac equation in the preceding sections; details on the semiclassical 
analysis of the Pauli equation with arbitrary spin can be found in 
\cite{KepPhd}. 

The Pauli equation for a spin-$\frac{1}{2}$ particle can be obtained from 
the Dirac equation \eqref{Dirac_eq} in the non-relativistic limit 
$c\to\infty$, see e.g. \cite{BjoDre64,Tha92}. With the representation 
\eqref{alpha_beta_def} of the matrices $\vecalph$ and $\beta$ and writing 
the Dirac spinor as $\Psi^T=(\psi^T,\chi^T)$ with 
$\psi,\chi \in L^2(\R^3) \otimes \C^2$ one finds the following equation for 
the upper two components, 
\begin{equation}
\label{Pauli_eq}
  \op{H}_\mathrm{P} \psi(\vecx) = E \psi(\vecx) 
\end{equation} 
with Pauli Hamiltonian 
\begin{equation}
\label{Pauli_Hamiltonian_1/2}
  \op{H}_\mathrm{P} 
  = -\frac{\hbar^2}{2m} 
    \left( \frac{\hbar}{\ui}\nabla - \frac{e}{c}\vecA(\vecx) \right)^2
    + e\phi(\vecx) - \frac{e}{mc} \vecB(\vecx) \, \frac{\hbar}{2} \vecsig \, .
\end{equation}
Let us generalise our discussion to Pauli Hamiltonians
\begin{equation}
\label{Pauli_Hamiltonian}
  \op{H}_\mathrm{P} 
  = \op{H}_\mathrm{S} + \frac{\hbar}{2} \ud\pi_s(\vecsig) \op{\vcB} \, , 
\end{equation}
where $\op{H}_\mathrm{S}$ is a Schr\"odinger Hamiltonian with Weyl symbol
$H(\vecp,\vecx)$, $\ud\pi_s(\vecsig)$ denotes the $2s+1$ dimensional (derived)
irreducible representation of $\su(2)$ and $\op{\vcB}$ is the Weyl quantisation
of the classical vector valued function $\vcB(\vecp,\vecx)$ on phase space.
The wave function $\psi$ has now $2s+1$  components, 
i.e $\psi \in L^2(\R^d) \otimes \C^{2s+1}$. The special case 
\eqref{Pauli_Hamiltonian_1/2} is recovered by the choices 
$d=3$, $s=\frac{1}{2}$, $H(\vecp,\vecx) 
= \frac{1}{2m} \left( \vecp -\frac{e}{c} \vecA(\vecx) \right)^2 + e\phi(\vecx)$
and $\vcB(\vecp,\vecx)=-\frac{e}{mc}\vecB(\vecx)$. Spin-orbit coupling 
can, e.g., be described by 
\begin{equation}
\label{spin-orbit-B}
  \vcB_\mathrm{so}(\vecp,\vecx) = f(r) \vecL \, , 
\end{equation}
where $\vecL=\vecx \times \vecp$ is orbital angular momentum and $f$ is an
arbitrary function of the radial coordinate $r=|\vecx|$. 

As in \eqref{WKBansatz} we make the ansatz 
\begin{equation}
\label{pauli_ansatz}
  \psi(\vecx) 
  = \sum_{k\geq0} \left( \frac{\hbar}{\ui} \right)^k a_k(\vecx) \, 
    \ue^{\frac{\ui}{\hbar}S(\vecx)}
\end{equation}
with scalar $S$, and $a_k$ taking values in $\C^{2s+1}$. Upon inserting 
\eqref{pauli_ansatz} into the Pauli equation \eqref{Pauli_eq} with Hamiltonian 
\eqref{Pauli_Hamiltonian}, in leading order we find the Hamilton-Jacobi
equation 
\begin{equation}
  H(\nab{x}S,\vecx) = E
\end{equation}
and in next-to-leading order we obtain the transport equation 
\begin{equation}
\label{pauli_transport}
  \left[ (\nab{p}H)(\nab{x}S,\vecx) \right](\nab{x}a_0) 
  + \frac{1}{2} \left[ \nab{x} (\nab{p}H)(\nab{x}S,\vecx) \right] a_0
  + \frac{\ui}{2} \ud\pi_s(\vecsig) \vcB(\nab{x}S,\vecx) \, a_0
  = 0 \, .
\end{equation}
As below \eqref{transport_b} the ansatz 
\begin{equation}
  a_0(\vecx) = \sqrt{\det\frac{\partial\vecy}{\partial\vecx}} \, u(\vecx)
\end{equation}
leaves us with the spin transport equation
\begin{equation}
  \dot{u}(\vecx) 
  + \frac{\ui}{2} \ud\pi_s(\vecsig) \vcB(\nab{x}S,\vecx) \, u(\vecx) = 0 \, .
\end{equation}
Integration along a trajectory with $\phi_H^t(\vecxi,\vecy) = (\vecp,\vecx)$
yields 
\begin{equation}
  u(\vecx) = \pi_s(d(\vecxi,\vecy,t)) \, u(\vecy) \, , 
\end{equation}
where $\pi_s$ denotes the $2s+1$ dimensional unitary irreducible 
representation of $\SU(2)$ and $d$ solves the spin transport equation 
\begin{equation}
  \dot{d}(\vecxi,\vecy,t) 
  + \frac{\ui}{2} \vecsig \vcB\!\left(\phi_H^t(\vecxi,\vecy)\right) 
    d(\vecxi,\vecy,t) = 0 \, , \quad 
  d(\vecxi,\vecy,0) = \eins_2 \, ,  
\end{equation}
in $\SU(2)$. Thus the covering map $\varphi: \SU(2) \to \SO(3)$ still relates
spin transport to classical spin precession,
\begin{equation}
  \dot{\vecs} = \vcB\!\left(\phi_H^t(\vecxi,\vecy)\right) \times \vecs
  \, , \quad \vecs \in S^2 \hookrightarrow \R^3 \, ,
\end{equation}
and we identify the skew product flow
\begin{equation}
  Y_\mathrm{cl}^t(\vecp,\vecx,\vecs) = 
  \left( \phi_H^t(\vecp,\vecx), \varphi(d(\vecp,\vecx,t)) \vecs \right)
\end{equation}
as the classical system corresponding to the Pauli equation with Hamiltonian 
\eqref{Pauli_Hamiltonian}. If $Y_\mathrm{cl}^t$ is integrable in the sense 
of Definition \ref{def_integrable_Ycl} we can use the same construction as 
in section \ref{sec:rotation_angels} to define a semiclassical wave 
function associated with a Liouville-Arnold torus $\T^d$. Locally we have
\begin{equation}
\label{sc_wave_function_Pauli}
  \psi_{\rm sc}(\vecx) \sim 
  \sqrt{\det\frac{\partial\vecy}{\partial\vecx}} \, 
  \ue^{\frac{\ui}{\hbar}S(\vecx)} \,
  \pi_s(d(\vecxi,\vecy,t)) \, u(\vecy) \, .
\end{equation}
and the initial and final values of $u$ after transport along along a basis
cycle $\mathcal{C}_j$ are related by $\pi_s(d_j(\vecxi,\vecy))$. As in 
\eqref{def_alpha_j} we obtain
\begin{equation}
  d_j(\vecxi,\vecy) 
  = \ue^{-\frac{\ui}{2}\alpha_j \vecsig\vecn(\vecxi,\vecy)} \, , 
\end{equation}
where the axis $\vecn$ is specified by Theorem \ref{integrable_Ycl} 
and the angle $\alpha_j$ is the rotation angle for a classical spin vector.
The representation matrix $\pi_s(d_j(\vecxi,\vecy))$ has eigenvalues 
$\exp(-\ui m_s \alpha_j)$, $m_s=-s,-s+1,\hdots,s$, see e.g. 
\cite{BarRac77}. Upon choosing $n(\vecxi,\vecy)$ to be an eigenvector of
$\pi_s(d_j(\vecxi,\vecy))$ the total phase change of experienced by
$\psi_\mathrm{sc}$ along a cycle $\mathcal{C}_j$ is given by 
\begin{equation}
  \frac{1}{\hbar} \oint_{\mathcal{C}_j} \vecP \, \ud\vecX 
  - \ui \frac{\pi}{2} \mu_j - m_s \alpha_j
\end{equation} 
thus resulting in the semiclassical quantisation conditions 
\begin{equation}
\label{spin_quant_pauli}
  \oint_{\mathcal{C}_j} \vecP \, \ud\vecX 
  = 2\pi\hbar \left( n_j + \frac{\mu_j}{4} + m_s \frac{\alpha_j}{2\pi} \right)
  , 
\end{equation}
with $\vecn \in \Z^d$ and $m_s=-s,-s+1,\hdots,s$.

\section{Spherically symmetric systems}
\label{sec:spherical}

Before we treat some explicit examples in order to illustrate the novel 
quantisation conditions \eqref{spin_quant} and \eqref{spin_quant_pauli}
we show how to apply them to an important class of integrable systems, 
namely spherically symmetric systems.

A spherically symmetric Dirac Hamiltonian has the structure
\begin{equation}
\label{spherical_Dirac}
  \op{H}_\mathrm{D} 
  = c\vecalph \, \frac{\hbar}{\ui}\nabla + \beta \, mc^2 + e\phi(r) \, , 
\end{equation}
where the electrostatic potential $\phi$ depends only on the radial variable
$r:=|\vecx|$. One easily verifies that the Hamiltonian \eqref{spherical_Dirac}
commutes with all components of total angular momentum, 
\begin{equation}
  \op{\vecJ}_\mathrm{D} := \op{\vecL} 
  + \frac{\hbar}{2} \begin{pmatrix} \vecsig & 0 \\ 0 & \vecsig \end{pmatrix}
  \, , \quad
  \op{\vecL} := \vecx \times \frac{\hbar}{\ui}\nabla \, .
\end{equation}
Since one also has 
\begin{equation}
  [ \op{\vecJ}_\mathrm{D}^2, \op{J}_{\mathrm{D}z}] = 0
\end{equation}
one can choose the eigenfunctions of $\op{H}_\mathrm{D}$ to be also
simultaneous eigenfunctions of $\op{\vecJ}_\mathrm{D}^2$ and 
$\op{J}_{\mathrm{D}z}$.

Analogously a spherically symmetric Pauli Hamiltonian is of the form 
\begin{equation}
\label{spherical_Pauli}
  \op{H}_\mathrm{P} 
  = -\frac{\hbar^2}{2m} \Delta + e \phi(\vecx) 
    + \frac{\hbar}{2} \ud\pi_s(\vecsig) \, f(r) \op{\vecL} \, , 
\end{equation}
cf. \eqref{spin-orbit-B}, and commutes with all components of 
total angular momentum
\begin{equation}
\label{J_P_Def}
  \op{J}_\mathrm{P} := \op{\vecL} + \frac{\hbar}{2} \ud\pi_s(\vecsig) \, .
\end{equation}
Again due to 
\begin{equation}
  [ \op{\vecJ}_\mathrm{P}^2, \op{J}_{\mathrm{P}z}] = 0
\end{equation}
the eigenfunctions of $\op{H}_\mathrm{P}$ can be chosen such that they are 
simultaneous eigenfunctions to the modulus squared $\op{\vecJ}_\mathrm{P}^2$
and the $z$-component $\op{J}_{\mathrm{P}z}$ of total angular momentum.

In the semiclassical limit the Hamiltonians \eqref{spherical_Dirac} and 
\eqref{spherical_Pauli} both give rise to skew products $Y_\mathrm{cl}^t$
on $\R^d \times \R^d \times S^2$ with a spherically symmetric classical 
Hamiltonian $H(\vecp,r)$ and spin precession with fields of the form 
\begin{equation}
\label{spherical_so-coupling}
  \vcB(\vecp,\vecx) = f(r) \, \vecL \, .
\end{equation}
It is well-known that $\phi_H^t$ is integrable in the sense of Liouville 
and Arnold since 
\begin{equation}
\label{HLM_Poisson}
  \{H, L\} = \{H, M\} = \{L, M\} = 0 
\end{equation}
where $L$ and $M$ are the modulus and $z$-component of classical orbital 
angular momentum $\vecL = \vecx \times \vecp$, respectively.
We remark that $L(\vecp,\vecx)$ not being smooth at points where 
$\vecp\|\vecx$ will not play a r\^ole in the follwing since all relevant 
constructions will stay away from these points. Thus \eqref{HLM_Poisson} 
holds wherever needed.

In order to show that $Y_\mathrm{cl}^t$ with field 
\eqref{spherical_so-coupling} is also integrable we have to extend the 
flows $\phi_L^t$ and $\phi_M^t$ to skew products with fields $\vcB_L$ and 
$\vcB_M$ fulfilling \eqref{spin_involution}. To this end consider 
the Weyl symbol of the operator 
$\op{\vecJ}:= \op{\vecL} + \frac{\hbar}{2} \vecsig$, 
\begin{equation}
  \vecJ = \vecp \times \vecx + \frac{\hbar}{2} \vecsig \, .
\end{equation}
By straightforward calculation one finds 
\begin{equation}
  J:= |\vecJ| = L + \frac{\hbar}{2} \vecsig \, \frac{\vecL}{L} + \O(\hbar^2)
  \quad \text{and} \quad 
  J_z = M + \frac{\hbar}{2} \sigma_z \, .
\end{equation}
Comparing these to the Pauli Hamiltonian \eqref{spherical_Pauli}
suggests that 
\begin{equation}
  \vcB_L := \frac{\vecL}{L} 
  \quad \text{and} \quad 
  \vcB_M := \vece_z \, , 
\end{equation}
where $\vece_z$ is the unit vector in $z$-direction, might be a good 
choice. As one easily checks the skew products
$Y_\mathrm{cl}^t$, $Y_L^t$ and $Y_M^t$ obtained in this way indeed form a set 
of commuting flows on $\R^d \times \R^d \times S^2$ and thus 
$Y_\mathrm{cl}^t$ is integrable, cf. Theorem \ref{integrable_Ycl}.

Spherically symmetric Hamiltonians $H(\vecp,r)$ can be separated in 
spherical coordinates ($r,\theta,\phi)$, see e.g. \cite{Gol80}. Introducing 
the action variables 
\begin{equation}
  I_r = \oint p_r \, \ud r \, , \quad
  I_\theta = \oint p_\theta \, \ud \theta 
  \quad \text{and} \quad
  I_\phi = \oint p_\phi \, \ud \phi \, , 
\end{equation}
where $p_r$, $p_\theta$ and $p_\phi$ denote the canonical momenta conjugate to 
$r$, $\theta$ and $\phi$, respectively, one always finds
\begin{equation}
  I_\phi = M 
  \quad \text{and} \quad 
  I_\theta = L - M \, .
\end{equation}
The radial action $I_r$ depends on the particular system under investigation 
and solving $I_r = \oint p_r(E,I_\theta,I_\phi) \ud r$ for $E$ yields the 
Hamiltonian $\overline{H}$ in action and angle variables. It is a general 
feature of spherically symmetric systems that $\overline{H}$ does 
not depend on $I_\theta$ and $I_\phi$ independently but only on the sum 
$L=I_\theta + I_\phi$, i.e. $\overline{H}$ is a function of $I_r$ and $L$.

The motion generated by $M=I_\phi$ is a rotation about the $z$-axis 
and thus there is no turning point in the time evolution of the coordinate 
$\phi$ yielding the Maslov index
\begin{equation}
 \mu_M = \mu_\phi = 0 \, .
\end{equation}
On the other hand the motion generated by $I_\theta$ takes place 
between two turning points for the coordinate $\theta$ and thus 
\begin{equation}
  \mu_\theta = 2 
  \quad \text{and accordingly} \quad
  \mu_L = 2 \, .
\end{equation}
In order to find the spin rotation angle $\alpha_M$ one has to integrate 
the spin precession equation 
\begin{equation}
  \dot{\vecs} = \vece_z \times \vecs
\end{equation}
over one cycle of $\phi_M^t$. The latter simply changes the angle variable 
$\vartheta_M$ conjugate to $M$ with unit speed and we have 
\begin{equation}
  \alpha_M = 2\pi|\vece_z| = 2\pi \, .
\end{equation}
Similarly $\phi_L^t$ changes $\vartheta_L$ with unit speed and since
$\vcB_L = \vecL/L$ is constant along such a cycle we also find
\begin{equation}
  \alpha_L = 2\pi|\vecL/L| = 2\pi \, .
\end{equation}
Therefore, for any spherically symmetric Dirac or Pauli equation we have 
semiclassical quantisation conditions
\begin{equation}
\label{quantise_L_M}
  L = \hbar \left( l + \frac{1}{2} + m_s \right)
  \quad \text{and} \quad
  M = \hbar \, (m_l+m_s) 
\end{equation}
with integers $l$ and $m_l$.

Classical mechanics imposes the restrictions 
\begin{equation}
\label{restrictLM}
  L \geq 0 
  \quad \text{and} \quad
  M \leq L \, .
\end{equation}
Defining the new quantum numbers $j:=l+m_s$ and $m_j:=m_l+m_s$ after some
trivial algebra one finds that \eqref{restrictLM} translates to
\begin{equation}
  j \geq 0 
  \quad \text{and} \quad
  m_j = -j, -j+1, \hdots, j \, .
\end{equation}
Notice that if $s$ is integer or half-integer, respectively, then so are 
$j$ and $m_j$. Finally the semiclassical quantisation conditions for angular 
momentum read 
\begin{equation}
\label{quantiseJ}
  L = \hbar \left( j + \frac{1}{2} \right)
  \quad \text{and} \quad
  M = \hbar \, m_j \, . 
\end{equation}

As mentioned at the end of section \ref{sec:rotation_angels} it will now be 
useful if we also find a field $\vcB_r$ which turns the flow $\phi_r^t$ 
generated by the action variable $I_r$ into a skew product that commutes with 
$Y_L^t$ and $Y_M^t$. To this end we may exploit the relation 
\eqref{consistent_B}. Keeping in mind that $\overline{H}$ is a function of 
$I_r$ and $L=I_\theta+I_\phi$ only, i.e. 
$\omega_\theta=\omega_\phi=\omega_L=\partial\overline{H}/\partial L$, 
this yields 
\begin{equation}
  \vcB_r = \frac{\vcB - \omega_L \vcB_L}{\omega_r} \, .
\end{equation}
Since $\vcB = f(r) \vecL$ is parallel to $\vcB_L = \vecL/L$ and since
$\vecL$ is a constant of motion $\vcB_r$ does not change its 
direction along the flow line of either $\phi_H^t$ or $\phi_r^t$. 
Therefore
\begin{equation}
  \alpha_r = \left| \int_0^{2\pi} 
  \frac{\vcB - \omega_L \frac{\vecL}{L}}{\omega_r} \ud\vartheta_r \right| ,
\end{equation}
where $\vartheta_r$ is the angle variable conjugate to $I_r$. Splitting the 
integrand into two terms, $\vcB/\omega_r$ and $-\omega_L \vecL/(\omega_r L)$,
the second integration is again trivial. In the first integral we can make  
use of $\ud\vartheta_r/\omega_r = \ud\vartheta_r/\dot{\vartheta}_r = \ud t$,
where $t$ is the physical time along a flow line of $\phi_H^t$. Therefore
we have obtained the handy expression 
\begin{equation}
\label{alpha_r_allg}
  \alpha_r = \left| \, \oint\limits_\mathrm{radial} 
  \vcB\!\left(\phi_H^t(\vecp,\vecx)\right) \ud t 
  - 2\pi\frac{\omega_L}{\omega_r} \frac{\vecL}{L} \right| 
\end{equation} 
where the remaining integral extends over one cycle of the radial motion, 
e.g. from perihelion to aphelion and back. The missing quantisation 
condition then reads 
\begin{equation}
\label{quantise_I_r}
  I_r = 2\pi\hbar \left( 
        n_r + \frac{1}{2} + m_s \frac{\alpha_r}{2\pi} \right) \, , 
\end{equation}
where we have inserted $\mu_r=2$ for a typical radial motion between 
perihelion and aphelion.

\section{Example 1:\\ Harmonic oscillator with spin-orbit coupling}
\label{sec:oscillator}

As a first example we discuss the three dimensional isotropic harmonic 
oscillator with spin-orbit coupling in a non-relativistic context with 
arbitrary spin. 

The Pauli Hamiltonian
\begin{equation}
\label{3DHOmSO_Def}
  \op{H}_\mathrm{P} 
  = -\frac{\hbar^2}{2m}\Delta + \frac{m}{2}\omega^2 r^2 
    + \frac{\hbar}{2} \ud\pi_s(\vecsig) \, \kappa \op{\vecL}
\end{equation}
describes an oscillator with frequency $\omega$ and for the spin-orbit 
coupling we have chosen a Thomas term, i.e. the leading order expression 
of $\vcB^+$, see \eqref{transport_b}, in the non-relativistic limit
$c\to\infty$ with 
\begin{equation}
  e\vecE = -\nabla \left( \frac{m}{2}\omega^2 r^2 \right)
         = -m\omega^2 \vecx
\end{equation}
i.e.
\begin{equation}
  \vecB^+ = \frac{\omega^2}{2mc^2}\vecL + \O(c^{-4}) \, .
\end{equation}
By comparison with \eqref{3DHOmSO_Def} we have $\kappa = \omega^2/(2mc^2)$
but we may keep $\kappa$ arbitrary for the rest of the section.

One usually determines the eigenvalues of \eqref{3DHOmSO_Def} 
by first observing that besides $\op{\vecJ}_\mathrm{P}^2$ and 
$\op{J}_{\mathrm{P}z}$, see \eqref{J_P_Def}, also $\op{\vecL}^2$ and 
$\op{\vecs}^2 = [\hbar/2 \ud\pi_s(\vecsig)]^2$ commute with 
$\op{H}_\mathrm{P}$ and that the spin-orbit term can be expressed in terms
of these conserved quantities, see e.g. \cite{Str98}.

Obviously, the Hamiltonian \eqref{3DHOmSO_Def} defines a spherically 
symmetric system as discussed in the preceeding section. Therefore, we 
immediately have the quantisation conditions \eqref{quantiseJ} for 
total angular momentum. The semiclassical eigenvalues, however, can 
even be found in a more straightforward way than to draw on 
\eqref{alpha_r_allg}. To this end notice that the classical Hamiltonian 
$H(\vecp,r) = \frac{\vecp^2}{2m} + \frac{m}{2}\omega^2r^2$ transformed
to action and angle variables reads
\begin{equation}
  \overline{H}(I_r,L) = \omega(2I_r+L) \, .
\end{equation}
Semiclassical quantisation with $s=0$ yields 
\begin{equation}
\label{3DHO_ohne_spin}
  E_{n_rl} = \hbar\omega \left( 2n + l + \frac{3}{2} \right) ,
\end{equation}
exhibiting the zero point energy of $3\hbar\omega/2$ of the three independent
oscillators in $x$-, $y$- and $z$-direction.
Since $\overline{H}$ only depends on a linear combination of $I_r$ and $L$
the problem has an even higher degeneracy than general spherically 
symmetric systems. Thus we may introduce the new action variable 
\begin{equation}
  I_1 := 2I_r + L
\end{equation}
with corresponding frequency 
$\omega_1 = \partial\overline{H}/\partial I_1 = \omega$. 
From \eqref{3DHO_ohne_spin} we see that without spin $I_1$ has to be quantised 
as $I_1 = \hbar(2n + l + \frac{3}{2})$. In order to find the correction from 
the spin contribution we have to extend $\phi_{I_1}^t$ to a skew product 
$Y_{I_1}^t$ that commutes with $Y_\mathrm{cl}^t$. The consistency condition 
\eqref{consistent_B} uniquely determines the relevant field $\vcB_1$, 
since it reduces to 
\begin{equation}
  \vcB = \omega_1 \vcB_1 
  \quad \Rightarrow \quad
  \vcB_1 = \frac{\vcB}{\omega_1} = \frac{\kappa}{\omega} \vecL \, .
\end{equation}
Integration of the spin precession equation 
$\dot{\vecs} = \vcB_1 \times \vecs$ is once more trivial 
since $\vecL$ is a constant of motion and we get 
\begin{equation}
  \alpha_1 = \frac{2\pi\kappa}{\omega} |\vecL| 
  = \frac{2\pi\hbar\kappa}{\omega} \left( l + \frac{1}{2} + m_s \right) \, .
\end{equation}
Finally, the semiclassical energies read
\begin{equation}
  E_{n_r l m_s} = \hbar\omega \left( 2n + l + \frac{3}{2} \right) 
  + m_s \hbar^2 \kappa \left( l + \frac{1}{2} + m_s \right) \, .
\end{equation}
For spin $s=\frac{1}{2}$ these are the exact eigenvalues of the 
Hamiltonian \eqref{3DHOmSO_Def} whereas for $s\neq\frac{1}{2}$ they 
are good approximations to the exact eigenvalues if $l$ is large, 
i.e. if the action variable $L$ is large compared to $\hbar$ as 
required for semiclassical approximations.

At this point a short remark concerning the practical application of 
conditions \eqref{spin_quant} or \eqref{spin_quant_pauli} is in order:
In general the spin rotation angels $\vecalph$ can depend on the action
variables $\vecI$. If the dependence is simple as in this example 
($\alpha_1$ depends on $L$ but not on $I_1$, $\alpha_L$ and $\alpha_M$ are 
constant) one can recursively apply \eqref{spin_quant_pauli} as was done 
above: First we quantised $L$ and then used the result when quantising 
$I_1$. In general, however, one may first have to solve the quantisation 
conditions \eqref{spin_quant_pauli} for the action variables $\vecI$
before inserting the result into the Hamiltonian $\overline{H}(\vecI)$.

\section{Example 2:\\ Sommerfeld's theory of fine structure revisited}
\label{sec:sommerfeld}

In this section we address the classic problem of Sommerfeld's theory
of fine structure which was already mentioned in the introduction as
a motivating example. 

To this end we have to consider the relativistic  Kepler problem with 
classical Hamiltonian 
\begin{equation}
\label{Kepler_Hamiltonian}
  H(\vecp,r) = - \frac{e^2}{r} + \sqrt{c^2\vecp^2 + m^2c^4} \, .
\end{equation}
Solutions of Hamilton's equations of motion corresponding to bound states 
are given by ``Rosettenbahnen'', ellipses with moving perihelia. 
Since (\ref{Kepler_Hamiltonian}) is spherically symmetric angular momentum 
is conserved and the motion takes place in a plane. Introducing 
polar coordinates in this plane the orbits can be expressed as 
\begin{equation}
\label{rosette}
  \frac{1}{r(\phi)} 
  = \underbrace{\frac{e^2 E}{c^2L^2-e^4}}_{:=C} 
    + \underbrace{\frac{\sqrt{c^2L^2E^2 + (c^2L^2-e^4) m^2c^4}}{c^2L^2-e^4}}
      _{:=A} \, 
    \cos \bigg( \underbrace{\frac{\sqrt{c^2L^2-e^4}}{cL}}_{:=\gamma}
                \, \phi \bigg)  ,
\end{equation}
where for later reference we have defined the constants $C$, $A$ and $\gamma$.
Detailed information on the classical mechanics of the relativistic Kepler 
problem can, e.g., be found in Sommerfeld's original article \cite{Som16}
or in his book \cite{Som69}. Transforming the Hamiltonian to action and 
angle variables yields
\begin{equation}
\label{H_relKepler_IL}
  \overline{H}(I_r,L) 
  = mc^2 \left[ 1 + \frac{e^4/c^2}{\left(I_r 
                + \sqrt{L^2-e^4/c^2}\right)^2} \right]^{-1/2} \, .
\end{equation}
Sommerfeld quantised the system using \eqref{epstein} by demanding  
\begin{equation}
\label{Sommerfeld_L_I_r}
  I_r = \hbar n_r 
  \quad \text{and} \quad
  L = \hbar l
\end{equation}
with integers $n_r$ and $l$. Classical mechanics imposes the restrictions 
$I_r \geq 0$ and $L\geq0$ yielding $n_r,l\geq0$. Moreover, Sommerfeld excluded
$l=0$ because in this case the electron would collide with the nucleus.
Thus he found the energy levels
\begin{equation}
\label{Sommerfeld_energies}
  E_{n_r l}^\mathrm{Sommerfeld} 
  = mc^2 \left[ 1 + \frac{\alpha_\mathrm{S}^2}{
    \left( n_r + \sqrt{l^2-\alpha_\mathrm{S}^2}\right)^2} \right]^{-1/2} 
  \, , \quad n_r \in \N_0 \, , \quad l \in \N \, , 
\end{equation}
where $\alpha_\mathrm{S}=e^2/(\hbar c)$ denotes Sommerfeld's fine structure 
constant.

When now quantising the relativistic Kepler problem with spin $\frac{1}{2}$ 
using conditions \eqref{spin_quant} first notice that due to spherical 
symmetry we have, cf. \eqref{quantise_L_M},
\begin{equation}
  L = \hbar \left( l + \frac{1}{2} + m_s \right) 
  \, , \quad m_s =\pm \frac{1}{2} \, .
\end{equation}
Since the additional conditions \eqref{restrictLM} do not allow the 
combination $l=0$, $m_s=-\frac{1}{2}$ (which would lead to a negative 
quantum number $j$) we see that $L/\hbar \in \N$, i.e. it assumes the
same values as in Sommerfeld's prescription \eqref{Sommerfeld_L_I_r}.

In order to quantise the radial action variable $I_r$ we also have to 
calculate the spin rotation angle $\alpha_r$ to which end we can use the
general formula \eqref{alpha_r_allg}. The field $\vcB$ occurring in the 
equation of spin precession can be obtained from $\vcB^+$, see 
\eqref{transport_b}, by inserting $\vecE = - \nabla(-e^2/r)$ yielding
\begin{equation}
  \vcB = \frac{e^2c^2}{\varepsilon(\varepsilon+mc^2)} \frac{1}{r^3} \, \vecL 
  \, . 
\end{equation}
where $\varepsilon$ was defined in equation \eqref{epsilon_def}. Without 
restriction we may choose $\vecL\|\vece_z$ for our calculation leading to
\begin{equation}
  \vecL = \frac{\varepsilon}{c^2} \, r^2 \, \dot{\phi} \, \vece_z 
  \quad \Rightarrow \quad 
  \vcB = \frac{e^2}{\varepsilon+mc^2} \, \frac{1}{r} \, 
         \dot{\phi} \, \vece_z \, .
\end{equation}
Using $E=H(\vecp,r)=-e^2/r + \varepsilon$, see \eqref{Kepler_Hamiltonian}, 
this expression further simplifies to 
\begin{equation}
\label{vcB_Kepler}
  \vcB = \frac{e^2}{(E+mc^2)r+e^2} \, \dot{\phi} \, \vece_z \, . 
\end{equation}
which is nicely adapted for use in \eqref{alpha_r_allg},
\begin{equation}
  \oint\limits_\mathrm{radial} \vcB \, \ud t 
  = \vece_z \oint\limits_\mathrm{radial} 
    \frac{e^2}{(E+mc^2)r(\phi)+e^2} \, \ud\phi
  = \vece_z \oint\limits_\mathrm{radial} \left[ 1 - 
    \frac{E+mc^2}{E+mc^2 + e^2 / r(\phi)} \right] \ud\phi \, .
\end{equation}
Changing variables to $\eta = \gamma\phi$ and inserting \eqref{rosette}
we get
\begin{equation}
  = \frac{\vece_z}{\gamma} \oint\limits_\mathrm{radial} \left[ 1 - 
    \frac{E+mc^2}{E+mc^2 + e^2 C + e^2 A \cos(\eta)} \right] \ud\phi
  = \frac{\vece_z}{\gamma} (1-2\pi\gamma) \, , 
\end{equation}
where the calculation of the last integral is tedious but elementary.
From \eqref{H_relKepler_IL} we easily determine the frequency ratio 
in the second term of \eqref{alpha_r_allg} to $\omega_L/\omega_r = \gamma^{-1}$
and since we have chosen $\vecL\|\vece_z$ we obtain
\begin{equation}
  \alpha_r = \left| \vece_z \left(\frac{1}{\gamma} -2\pi \right) 
                    - \frac{\vece_z}{\gamma} \right| 
           = 2\pi \, .
\end{equation}
This remarkable result turns \eqref{quantise_I_r} into the rather trivial
condition 
\begin{equation}
  I_r = \hbar \left( n_r + \frac{1}{2} + m_s \right)
\end{equation}
with integer $n_r$ and $m_s=\pm\frac{1}{2}$. Moreover, classical mechanics 
demands $I_r\geq0$ and thus $I_r/\hbar \in \N_0$, i.e. the values assumed
are again identical to those predicted by Sommerfeld's condition 
\eqref{Sommerfeld_L_I_r}.

Finally we have found the semiclassical energies,
\begin{equation}
\label{fine_structure}
  E_{n_r l m_s} = mc^2 \left[ 1 + \frac{\alpha_\mathrm{S}^2}{
    \left( n_r + \frac{1}{2} + m_s 
           + \sqrt{(l+\frac{1}{2}+m_s)^2-\alpha_\mathrm{S}^2}\right)^2} 
  \right]^{-1/2} \, , 
\end{equation}
which are identical to those obtained from the exact solution of the Dirac 
equation \eqref{Dirac_eq} with potential $-e^2/r$, see e.g. 
\cite{BjoDre64,Str98}, as can be most easily checked by introducing the 
quantum numbers $j$ and $m_j$ of total angular momentum, see below 
\eqref{restrictLM}, and the principal quantum number $n$, associated with 
$I := I_r + L$, by $I=\hbar n$. Although Sommerfeld's formula 
\eqref{Sommerfeld_energies} also yields the correct energy levels of 
the hydrogen atom it predicts the wrong multiplicities. This problem is 
rectified by the present treatment as can be checked straightforwardly.
Consider, e.g., the ground state which in \eqref{Sommerfeld_energies} is 
obtained by $n_r=0$, $l=1$, and thus is non-degenerate. On the other hand 
in \eqref{fine_structure} the same energy is obtained by choosing either 
$n_r=-1$, $l=0$, $m_s=\frac{1}{2}$, or $n_r=0$, $l=1$, $m_s=-\frac{1}{2}$.
Alternatively these states can be characterised by $n=1$, $j=\frac{1}{2}$,
$m_j=\pm\frac{1}{2}$ yielding a multiplicity of 2 as in the exact 
quantum spectrum.

\section{Conclusions}
\label{sec:conclusions}

In this article 
we have derived semiclassical quantisation conditions for the Dirac 
and Pauli equations. We have shown that spin yields a contribution of the 
same order of magnitude as the Maslov correction. The spin contribution
is determined by rotation angles for a classical spin vector that is 
transported along orbits of the translational dynamics. 

The crucial step in the derivation of the semiclassical quantisation 
conditions was the generalisation of the notion of integrability to 
certain skew products and group extensions. The relevant integrability 
conditions enabled us to effectively reduce the non-Abelian Berry phases,
appearing in the analysis of multi-component wave equations, to a
$\U(1)$-holonomy, i.e. to an ordinary phase factor. The latter can then be
incorporated into the quantisation conditions.

We remark that our treatment generalises to arbitrary multi-component 
wave equations for which the principal symbol of the Hamiltonian has
eigenvalues with arbitrary but constant multiplicity. One then has to
deal with $G$-extensions, where $G=\U(n)$ or a subgroup thereof, 
of the ray dynamics (generated by the eigenvalues of the principal symbol). 
Proposition \ref{Gext_int} provides us with the relevant integrability 
condition which allows for an effective reduction of the holonomy group 
$G$ to an Abelian subgroup $H$. The spin rotation angles are then replaced by 
the eigenphases of some unitary representation matrices of $H$.

In section \ref{sec:sommerfeld} we have applied the novel 
quantisation conditions to the relativistic Kepler problem. We have seen 
that by a freak of nature all relevant spin rotation angles are given by 
$2\pi$ and thus cancel (or add up to an integer) with the Maslov term.
It is this coincidence due to which Sommerfeld was able to calculate 
the energy levels of the relativistic hydrogen atom including spin-orbit
coupling 10 years before the Dirac equation was developed.

\subsection*{Acknowledgement}

I would like to thank Jens~Bolte for helpful discussions.
This work was supported by the Deutsche Forschungsgemeinschaft under contract 
no. Ste 241/10-2.

\begin{appendix}
\section{Wigner-Weyl calculus}
\label{weyl}

With a differential operator $\op{A}$ one can associate an object 
on classical phase space, its Weyl symbol $A(\vecp,\vecx)$, by
\begin{equation}
\label{WeylaufPsi}
  (\op{A} \Psi)(\vecx) = \frac{1}{(2\pi\hbar)^d} \int_{\R^d} \int_{\R^d}
  A \left( \vecp, \frac{\vecx + \vecz}{2} \right) \,
  \ue^{\frac{\ui}{\hbar} \vecp (\vecx - \vecz)} \, \Psi(\vecz) \, 
  \ud^dz \, \ud^dp \, . 
\end{equation}
If $\Psi$ is a multi-component object, e.g. 
$\Psi \in L^2(\R^d) \otimes \C^{2s+1}$, then  
$A(\vecp,\vecx)$ is matrix valued. Reverting this reasoning, 
one can also associate an operator $\op{A}$ with a more general symbol 
$A(\vecp,\vecx)$, which does not necessarily 
correspond to a differential operator, via
(\ref{WeylaufPsi}). This procedure is known as Weyl quantisation 
and certain properties of symbols translate to properties of the operators,
thus leading to so-called pseudo-differential operators, see e.g. 
\cite{Fol89} for an introduction. 

If an operator $\op{A}$ can be represented by an integral kernel 
$K_A(\vecx,\vecy)$, i.e. 
\begin{equation}
  (\op{A} \Psi)(\vecx) = \int_{\R^d} K_A(\vecx,\vecy) \, \Psi(\vecy) \, \ud^dy 
  \, ,
\end{equation}
one obtains its Weyl symbol from 
\begin{equation}
  A(\vecp,\vecx) = \int_{\R^d} K_A\!\left( \vecx + \frac{\vecz}{2} , 
                                          \vecx - \frac{\vecz}{2} \right)
                   \ue^{- \frac{\ui}{\hbar} \vecz \vecp} \, \ud^dz \, .
\end{equation}
Inverting this transformation yields 
\begin{equation}
  K_A(\vecx,\vecy) = \frac{1}{(2\pi\hbar)^d} \int_{\R^d} 
    A\!\left( \vecp, \frac{\vecx+\vecy}{2} \right) 
    \ue^{\frac{\ui}{\hbar}\vecp(\vecx-\vecy)} \, \ud^d p \, . 
\end{equation}
If the symbol $A(\vecp,\vecx)$ has an expansion in powers of $\hbar$,
\begin{equation}
\label{semiclassical_operator}
  A(\vecp,\vecx) = \sum_{k\geq0} \hbar^k A_k(\vecp,\vecx) \, , 
\end{equation}
the corresponding operator $\op{A}$ is called a semiclassical Weyl operator.
The leading and subleading terms $A_0$ and $A_1$ in the expansion 
\eqref{semiclassical_operator} are known as the principal symbol and the
subprincipal symbol, respectively.

Application of a semiclassical Weyl operator to a rapidly oscillating function 
\begin{equation}
\label{rapid_osc}
  \Psi_{\rm sc}(\vecx) = a_{\hbar}(\vecx) \, \ue^{\frac{\ui}{\hbar} S(\vecx)}
  \ , \quad
  a_{\hbar}(\vecx) 
  = \sum_{k=0}^{\infty} \left( \frac{\hbar}{\ui} \right)^k \, a_k(\vecx) \, .
\end{equation}
is governed by the following theorem; the corresponding statement in a
slightly different setting can, e.g., be found in \cite[chapter 4.3]{Dui96}.

\begin{theorem}
\label{theorem:WeylaufWKB}
Applying a semiclassical Weyl operator $\op{A}$ with a symbol $A(\vecp,\vecx)$
of the form \eqref{semiclassical_operator}
to a wave function of type \eqref{rapid_osc} yields in leading orders as 
$\hbar \to 0$, 
\begin{equation}
\begin{split}
\label{WeylaufWKB}
  (\op{A} \Psi_{\rm sc})(\vecx) &=  \bigg\{ 
  A_0(\nab{\vecx} S(\vecx), \vecx) \, a_0(\vecx)
  + \frac{\hbar}{{\rm i}} \bigg[
    A_0(\nab{\vecx} S(\vecx), \vecx) \, a_1(\vecx)
  \\ & \qquad
    + (\nab{\vecp} A_0)(\nab{\vecx} S(\vecx), \vecx) \, 
      \nab{\vecx} a_0(\vecx)
    + \frac{1}{2} a_0(\vecx) [\nab{\vecx} 
      (\nab{\vecp} A_0)(\nab{\vecx} S(\vecx),\vecx)] 
  \\ & \qquad
    + A_1(\nab{\vecx} S(\vecx), \vecx) \, a_0(\vecx)
  \bigg] + {\cal O}(\hbar^2) \bigg\} \, 
  \ue^{\frac{{\rm i}}{\hbar} S(\vecx)} \, .
\end{split}
\end{equation}
\end{theorem}
\noindent
Notice that the nesting of brackets in the fourth term on the r.h.s. 
indicates that the gradient with respect to $\vecp$ is taken before
we set $\vecp=\nab{x}S$, whereas the gradient with respect to $\vecx$
is only taken after doing so, i.e.
\begin{equation}
  [\nab{\vecx} (\nab{\vecp} A_0) (\nab{\vecx} S(\vecx),\vecx)] 
  = \sum_{j=1}^d \frac{\partial^2 A_0}{\partial p_j \partial x_j} 
                 (\nab{x}S(\vecx),\vecx)
    + \sum_{j=1}^d \sum_{j=1}^d  
      \frac{\partial^2 A_0}{\partial p_j \partial p_k}(\nab{x}S(\vecx),\vecx)
      \frac{\partial^2 S}{\partial x_j \partial x_k}(\vecx) \, .
\end{equation}
For the proof we refer the reader to \cite[chapter 4.3]{Dui96}.


\end{appendix}

\bibliographystyle{my_unsrt}              
\bibliography{literatur}                  

\end{document}